\documentclass{sigchi}

\CopyrightYear{2020}
\setcopyright{acmlicensed}
\doi{https://doi.org/10.1145/3313831.3376399}
\isbn{978-1-4503-6708-0/20/04}
\conferenceinfo{CHI'20,}{April  25--30, 2020, Honolulu, HI, USA}
\acmPrice{\$15.00}

\toappear{\scriptsize Permission to make digital or hard copies of all or part of this work for personal or classroom use is granted without fee provided that copies are not made or distributed for profit or commercial advantage and that copies bear this notice and the full citation on the first page. Copyrights for components of this work owned by others than the author(s) must be honored. Abstracting with credit is permitted. To copy otherwise, or republish, to post on servers or to redistribute to lists, requires prior specific permission and/or a fee. Request permissions from permissions@acm.org. \\
{\emph{CHI '20, April 25--30, 2020, Honolulu, HI, USA.} } \\
\copyright~2020 Copyright is held by the owner/author(s). Publication rights licensed to ACM. \\
ACM ISBN 978-1-4503-6708-0/20/04\ ...\$15.00.\\
\url{http://dx.doi.org/10.1145/3313831.3376399}}

\usepackage{balance}       
\usepackage{graphics}      
\usepackage[T1]{fontenc}   
\usepackage{txfonts}
\usepackage{mathptmx}
\usepackage[pdflang={en-US},pdftex]{hyperref}
\usepackage{color}
\usepackage{booktabs}
\usepackage{textcomp}
\usepackage{xspace}

\usepackage{microtype}        
\usepackage{ccicons}          

\usepackage{todonotes}

\newcommand{\harish}[1]{{\color{green} Harish: [{#1}]}}
\newcommand{\maryam}[1]{}

\newcommand{\highlight}[1]{{{#1}}}

\def\plaintitle{Urban Mosaic: Visual Exploration of\\Streetscapes Using Large-Scale Image Data}
\def\plainauthor{Fabio Miranda, Marcos Lage, Harish Doraiswamy,
   Maryam Hosseini, Graham Dove, Claudio T. Silva}
\def\plainkeywords{Urban planning; Interactive visualization; Data analysis; Urban data}

\makeatletter
\def\url@leostyle{%
  \@ifundefined{selectfont}{
    \def\UrlFont{\sf}
  }{
    \def\UrlFont{\small\bf\ttfamily}
  }}
\makeatother
\def\pprw{8.5in}
\def\pprh{11in}

\definecolor{linkColor}{RGB}{6,125,233}
\hypersetup{%
  pdftitle={\plaintitle},
  pdfauthor={\plainauthor},
  pdfkeywords={\plainkeywords},
  pdfdisplaydoctitle=true, 
  bookmarksnumbered,
  pdfstartview={FitH},
  colorlinks,
  citecolor=black,
  filecolor=black,
  linkcolor=black,
  urlcolor=linkColor,
  breaklinks=true,
  hypertexnames=false
}

\newcommand{\Rspace} {{\mathbb R}}

\newcommand{\myparagraph}[1]{\noindent\textbf{#1}\xspace}
\newcommand{\hidecomment}[1]{}

\newcommand{\eg}{e.g.,\xspace}

\newcommand{\tool}{Urban~Mosaic\xspace}

\begin{document}

\title{\plaintitle}

\numberofauthors{1}
\author{%
  \alignauthor{Fabio~Miranda\textsuperscript{1}, Maryam~Hosseini\textsuperscript{2}, Marcos~Lage\textsuperscript{3}, Harish~Doraiswamy\textsuperscript{1},\\Graham~Dove\textsuperscript{1}, Cl\'audio~T.~Silva\textsuperscript{1}\\
 \affaddr{\textsuperscript{1}New York University; \textsuperscript{2}Rutgers University}; \textsuperscript{3}Universidade Federal Fluminense\\
 \email{\textsuperscript{1}\{fmiranda,harishd,grahamdove,csilva\}@nyu.edu}; \textsuperscript{2}mary.hosseini@rutgers.edu; \textsuperscript{3}mlage@ic.uff.br} \\
}

\maketitle

\begin{abstract}
Urban planning is increasingly data driven, yet the challenge of designing with data at a city scale and remaining sensitive to the impact at a human scale is as important today as it was for Jane Jacobs. We address this challenge with Urban Mosaic, a tool for exploring the urban fabric through a spatially and temporally dense data set of 7.7 million street-level images from New York City, captured over the period of a year.  Working in collaboration with professional practitioners, we use Urban Mosaic to investigate questions of accessibility and mobility, and preservation and retrofitting. In doing so, we demonstrate how tools such as this might provide a bridge between the city and the street, by supporting activities such as visual comparison of geographically distant neighborhoods, and temporal analysis of unfolding urban development. 
\end{abstract}

\ccsdesc[500]{Human-centered computing~Human computer interaction (HCI)}
\ccsdesc[300]{Human-centered computing~Visual analytics}
\ccsdesc[300]{Human-centered computing~Visualization toolkits}

\keywords{\plainkeywords}

\printccsdesc

\noindent\textit{"A sense of place is built up, in the end, from many little things too, some so small people take them for granted, and yet the lack of them takes the flavor out of the city."
\begin{flushright}
\centering (Jane Jacobs, Downtown is for People)
\end{flushright}
}

\section{Introduction}

For those of us living in or visiting the world's major cities, their dynamism and complexity are immediately apparent. Yet urban planners and designers must work in a context where any single intervention, perhaps aimed at altering just one aspect, can have a wide ranging impact on a variety of interrelated components~\cite{batty2009cities}, affecting things at both a macro \textit{city scale} and a micro \textit{human scale}. Examples of changes at a city scale include, suburbanization, economic deconcentration, modification to transport infrastructure, rezoning and/or gentrification of neighborhoods, and major renewal projects. Examples of changes at a human scale, on the other hand, are reflected in the city's urban fabric, and include aspects that make a city livable, encourage walking, and contribute to the perception of safety; in other words affect the day to day lives of its inhabitants. This might be manifested in lighting, shadow, sky exposure, open-front shops, the details of building facades,~etc.~\cite{gehl2013cities}.

Responding to these challenges at the city scale, planners and designers have been aided by a rapid growth in data from urban environments, and so are able to turn to computational methods and large-scale data analysis, which increase understanding by quantifying different aspects of the city (e.g.,~\cite{hammon2015data,batty2013big,becker2011tale, ferreira2013visual,quercia2014mining, psyllidis2015platform, lecue2014star}). However, while sensitivity to the impact of change at a human scale remains as important today as it was for Jane Jacobs and others in the 1950s and 1960s~\cite{laurence2006death}, analyses of suitable data, which emphasize qualitative, visual details, are often difficult and time-consuming to perform. Although there is an increasing availability of street-view images, which support a degree of virtual assessment and auditing of the built environment, their distribution is often temporally sparse and so analysis is limited.

This paper introduces \tool, a tool for visually exploring the urban fabric. It responds to the challenges practitioners face by employing a newly available spatially and temporally dense data set of street-level images from New~York~City~(NYC) 
\tool is a visual exploration system designed to help practitioners in urban planning and design gain insight into the human scale impact of changes in the urban fabric. It utilizes state-of-the-art computer vision techniques for image similarity search and clustering, together with efficient spatio-temporal selection and aggregation over the image metadata to visually explore and map this image data set. It further allows the analysis to be augmented using spatio-temporal urban data from a variety of other sources (e.g., census, transport, crime, weather, housing market, zoning, noise complaints). The image data set used in this work contains 7.7~million images captured in the Manhattan and Brooklyn boroughs between April 2016 and April 2017 using car-mounted cameras.
\tool has been developed as a collaboration between researchers in urban planning, visual analytics, and HCI. 
\highlight{%
%
%
We include a detailed description of system features as a way to offer necessary background to our approach.
In order to better understand the needs and potential use cases of practitioners, we undertook a requirements gathering process that included: representatives of Kohn~Pedersen~Fox~(KPF), a major architectural practice specializing in data-driven urban design and urban data analytics; Draw~Brooklyn, a practice run by the former Chief~Urban~Designer of NYC; and an occupational therapist specializing in urban accessibility for older people. We refer to our interactions with these practitioners, which include formative evaluation sessions, throughout the paper to offer the perspective of potential future users.
}

\myparagraph{Contributions.}
The work reported in this paper has both practical and theoretical implications, and contributes to HCI knowledge in two ways. From a theoretical perspective, we demonstrate the potential for images from large-scale street-view data sets to help bridge data-driven urban planning and design at the city scale and at the human scale. From a practical perspective, we present a tool that enables professionals to combine quantitative spatio-temporal urban data sets with qualitative spatio-temporal data sets of urban images, and use these to solve the real problems they identified.
\section{Background}
The built environment's impact on public health, social well-being and quality of life has been demonstrated with regard to levels of physical activity~\cite{doi:10.1177/0042098008093386, doi:10.1177/1524839903260595}, social inclusion \cite{thornton2016disparities, bise2018sidewalks}, social capital~\cite{rogers2013social}, willingness to walk~\cite{doi:10.1123/jpah.2018-0476}, perceptions of safety \cite{ariffin2013perceptions}, pedestrian fatalities \cite{doi:10.2105/AJPH.93.9.1456}, impact on pedestrians with special needs \cite{10.2307/26203303}, and risk of cardiovascular or respiratory diseases~\cite{sallis2012role, DiezRoux2003}. Features such as wide sidewalks, greenery, townhouse stoops, street-level activity, and ground floor properties with detail and windows, all contribute to a vibrant street life that attracts more people to public spaces and increases safety. This was Jane Jacobs' now famous ``eyes on the streets'' theory \cite{jacobs_death_1961}, which has greatly influenced our understanding of how the built environment at a human scale affects crime~\cite{jeffery1977crime, newman1972defensible} and perceptions of safety~\cite{ariffin2013perceptions}. Safety from accidents, such as tripping or falling, caused by potholes, obstructions, bad lighting, and slippery surfaces (due either to weather conditions or surface material), is also a key concern, particularly for older adults, pregnant women and people with disabilities~\cite{AGHAABBASI2018475}. Analysis of the built environment's impact on these public health and quality of life issues typically involves time-consuming and costly in-field auditing, which requires trained auditors to be present at the site in question to make assessments and record their observations based on auditing protocols~\cite{BROWNSON2009S99, harvey2014measuring, CLIFTON200795, marshall2010effect}. 

While a rapid growth in urban data gathering has resulted in the development of new tools and practices that offer urban planners and designers significant opportunities and benefits at the city scale (e.g.,~\cite{ hammon2015data,batty2013big,ferreira2013visual,quercia2014mining, reades2007cellular}), there have been far fewer such developments at the human scale. Efforts to quantify and assess the visual appearance of the built environment have typically been limited in terms of the number and area of locations that can be covered, since they mainly rely on in-field data collection and assessment procedures~\cite{rundle2011using, PURCIEL2009457}. Recently however, the advent of new computer vision algorithms has made it possible to use image data to measure the characteristics of the built environment~\cite{glaeser2018big, naik2014streetscore}, and the introduction of Google Street View (GSV)~ \cite{anguelov2010google} and Microsoft Streetside~\cite{kopf2010street}, which provide free panoramic images captured by specifically designed vehicle-mounted cameras, has offered new data sources. This has enabled researchers to tackle questions in urban planning, urban design, urban sociology, criminology and public health from a new perspective ~\cite{6875954, doersch:hal-01053876, naik2014streetscore, 10.1257/aer.p20161030}. GSV in particular has made virtual auditing possible~\cite{rundle2011using, badland2009understanding, badland2010can, kelly2012using, charreire2014using}, expanding and diversifying geographical coverage, and reducing time and labor requirements. However, while trained experts can now use GSV in assessments, they must manually explore whole collections of images to identify features of interest~\cite{glaeser2018big}. 
An additional layer of difficulty is added when satisfaction criteria are based on external data; for example, identifying pavements that might pose a danger to pedestrians only under certain weather conditions, or assessing the impact of construction projects. GSV now includes a timeline feature which allows users to view temporal change. However, data are often very sparse in this regard and so not typically suitable for monitoring the evolution of a location of interest, and therefore reducing the usefulness of GSV images. A temporally dense collection of images on the other hand makes it possible to visualize not only the different blocks, neighborhoods, and boroughs of a city, but also changes in their physical appearance over extended periods. 

This research was facilitated, and in part motivated, by the availability of a new data set comprising of images captured in NYC over the course of a year\maryam{motivated by the data (1)}. The data has been generated by cameras mounted on top of regular vehicles, and the images were captured at regular intervals as the cars travel throughout the city.
%
%
This meant that photographs could be taken continuously and, since these vehicles have been on the streets throughout the year, often repeating locations, it has resulted in a data set that is temporally dense\maryam{dense(2)}, covering not only different times of the day but various seasons as well. However, without specialized hardware, and with no human intervention, there is also no guarantee that consecutive photographs were taken to uniformly cover the streets, or that photographs were taken with the correct focus.
Yet having such a dense data set creates the opportunity to conduct temporal analyses and comparison of architectural features across geographically distant locations. Because of this it allows for assessment of the urban fabric at a scale not previously practical.
\section{Related Work}
Concern for planning and designing the urban built environment, and supporting the practice of professionals working in these fields, has been relatively limited at CHI. Prior work includes Underkoffler and Ishi \cite{underkoffler1999urp}, who present a luminous-tangible workbench as a system supporting urban planning,
and White and Feiner \cite{white2009sitelens} who offer augmented reality support for site visits, by visualizing relevant data. More recently, Mahyar et al. \cite{mahyar2018communitycrit} introduced a system for engaging communities
in elaborating and evaluating urban design ideas through micro-activities; and Pang et al. \cite{Pang:2019:CED:3290605.3300571} describe employing a 
location-based game to investigate transit commuters' use of community information. However, recent work by Saha~et~al.~\cite{Saha:2019:PSW:3290605.3300292}, introducing ``Project Sidewalk", is of particular relevance to the research we present here. Project Sidewalk a is web-based tool that takes a crowd-sourcing approach to virtual assessment of urban built environments, with both paid crowd-workers and volunteers identifying and labeling accessibility issues. Among the key issues they raise, regarding the ongoing viability of their approach, are the age of GSV panoramas and the quality and age of labels provided. 

However, the increasing availability of urban data sets creates opportunities to analyze and visualize cities in new ways. This has resulted in a number of visual analytics systems designed to interactively explore and analyze data ~\cite{ferreira2013visual,7390069,Doraiswamy:2018:IVE:3183713.3193559}, and seek insight into transportation and mobility~\cite{4677356,6876029}, air pollution~\cite{4376168}, real-estate ownership~\cite{Hoang-Vu:2014:TUR:2630729.2630746}, and shadow impact on public spaces~\cite{8283638}. A comprehensive survey of research into urban visual analytics is presented in Zheng~et~al.~\cite{7506246}. One area of new activity that can be revealing is geo-tagged social network data. Urban~Pulse~\cite{miranda2016urban} uses computational topology techniques and data from Twitter and Flickr to visualize spatio-temporal activity across various resolutions, and Urban~Space~Explorer~\cite{8047432} proposes an approach to explore public-space-related activity. Another is using computer vision algorithms to assess the built environment through street-view images, for example to: assess and map greenery and openness in urban settings \cite{li2015assessing,li2017mapping}, quantify the daily exposure of urban residents to eye-level street greenery~\cite{ye2018measuring}, extract land use information~\cite{li2016urban}, measure visual quality~\cite{tang2018measuring}, visual enclosure~\cite{yin2016measuring}, urban form~\cite{shen2018streetvizor} and sky exposure~\cite{carrasco2015using} of street spaces, and assess traffic signs~\cite{balali2015detection}, curb ramps~\cite{hara2014tohme} and urban landmarks~\cite{lander2017inferring}. Street-level images have also been used to predict relationships between a city's built environment and socioeconomic conditions~\cite{6875954}, and as the basis for automatically extracting a city's most distinctive visual elements~\cite{doersch:hal-01053876}.

Querying large-scale image data sets is a key element of the computer vision techniques that enable these new strands of research. Prior to 2012 this was largely based on image features extracted by local descriptors such as SIFT~\cite{Lowe:2004:DIF:993451.996342} or HOG~\cite{1467360}. However, recent work focuses on feature extraction through convolutional neural network~(CNN); for example, pre-trained on image data sets (e.g., ImageNet)~\cite{simonyan2014very}, based on a fine-tuned CNN model~\cite{babenko2014neural}, or using the feature representation extracted from an intermediate layer of a pre-trained CNN~\cite{sharif2014cnn,tolias2016particular}. See~\cite{zhou2017recent} for a survey of content-based image retrieval. We build on this prior research to show how geographically and temporally dense street-level image data can support the work of professional practitioners in urban planning, design and related disciplines. To do this we present a novel system that enables users to explore a first of its kind collection of images by visually composing contextual queries and searching for visually similar images. We enhance this with support for spatiotemporal analysis of these images in conjunction with other urban data sets.

\section{Introducing Urban Mosaic}

\tool, is a novel system for gaining insight into the human-scale impact of city-wide urban development through exploration of a temporally and geographically dense image data set\maryam{dense(4) }, utilizing spatio-temporal selection and aggregation, image similarity, and clustering queries, which can be augmented with additional spatio-temporal urban data.
%
First we describe the street-level image data set that motivated our work \maryam{motivated by data (2)} and underpins our discussion of \tool in this paper. We then discuss its system architecture, our use of deep convolutional neural networks to extract feature embeddings, our methods for efficiently computing image similarity, and the \tool user interface. %
\highlight{\tool is currently a working prototype, which we are using to probe the potential needs and wants of professional practitioners, and to explore alternative use cases.}

\subsection{Motivation}

\highlight{
Our work is motivated by the emergence of new image data sets that offer spatially and temporally dense representations of cities, and their potential for aiding urban design and planning practitioners in auditing and assessing urban environments. \maryam{I removed the systematic assessment sentence and merge it with the next} 
Current assessment practices involve in-field auditing which requires visiting neighborhoods on foot, taking pictures, most often on just a single day.
In addition to the cost of training and hiring auditors to carry out this work, it is time-consuming and can involve extensive travel~\cite{emery2003reliability}. Weather conditions and potential safety concerns~\cite{zhou2017recent} place further constraints on the capacity to undertake frequent or city-wide audits. This can mean important accessibility problems, such as the potentially hazardous condition of a damaged sidewalk after heavy rain, or under snow in winter, may not be captured during auditing. 

Moving beyond the basic logistical challenges of undertaking an in-field audit, practitioners must also find common elements across these large collections of images; functionality that is typically not present even in virtual auditing tools. For example, an urban designer might wish to identify specific architectural articulations, e.g., a certain column type or window design that signifies architectural style, and indicates important factors such as estimated year of construction and building materials. Extracting this level of detail is costly and time consuming, particularly when undertaken through in-field auditing across the city, but also when attempted manually through exploring collections such as GSV. It can also be crucial to audit and assess urban environments for safety on an ongoing basis. For example, to know the condition of curb cuts or crossings under different weather conditions. If not properly maintained, they can pose a potentially life threatening hazard, specifically to the more vulnerable groups such as seniors or people with movement or vision impairment. 

In our work on Urban Mosaic, we aim to facilitate the early stages of assessing and auditing the built environment, enabling practitioners to combine quantitative and qualitative data and perform fast queries of visual features over a large, spatially and temporally dense collection of street-level images\maryam{dense(6)}. Urban Mosaic can provide earlier insights into how the city might be experienced, enabling visual comparison of geographically distant areas, and identifying similarity by architectural features and urban data sets, such as census data. Urban Mosaic will not replace the need to visit neighborhoods and experience them in person, but it can provide insight to help filter those areas that need visiting, and build a richer picture to augment those neighborhoods that are audited.
}

\subsection{Data Set of Street-Level Images}
The street-level image data used in this research was gracefully provided by Carmera. The data has been generated by cameras mounted on vehicles. These cameras, each orthogonal to the other and facing a particular direction, capture images at regular intervals as the cars travel throughout NYC.
Unlike GSV, where cars are deployed specifically to capture street-level images, these are vehicles making journeys as part of the driver's regular day-to-day activity. Because of this, the quality of images may vary due to illumination, weather, traffic conditions, or blurriness (e.g., due to vehicular speed). The mobile phone also records metadata for each image in the data set: time, location, and camera orientation. This data is more temporally dense\maryam{dense(7)} than data sets such as GSV or Microsoft Streetside. We employ a subset of the full data, covering the boroughs of Manhattan and Brooklyn between April 2016 to April 2017, and totaling \textit{7.77~million images}. \maryam{ the exact description in the intro. data size(2)}

\begin{figure}[t]
	\centering
	\includegraphics[width=\linewidth]{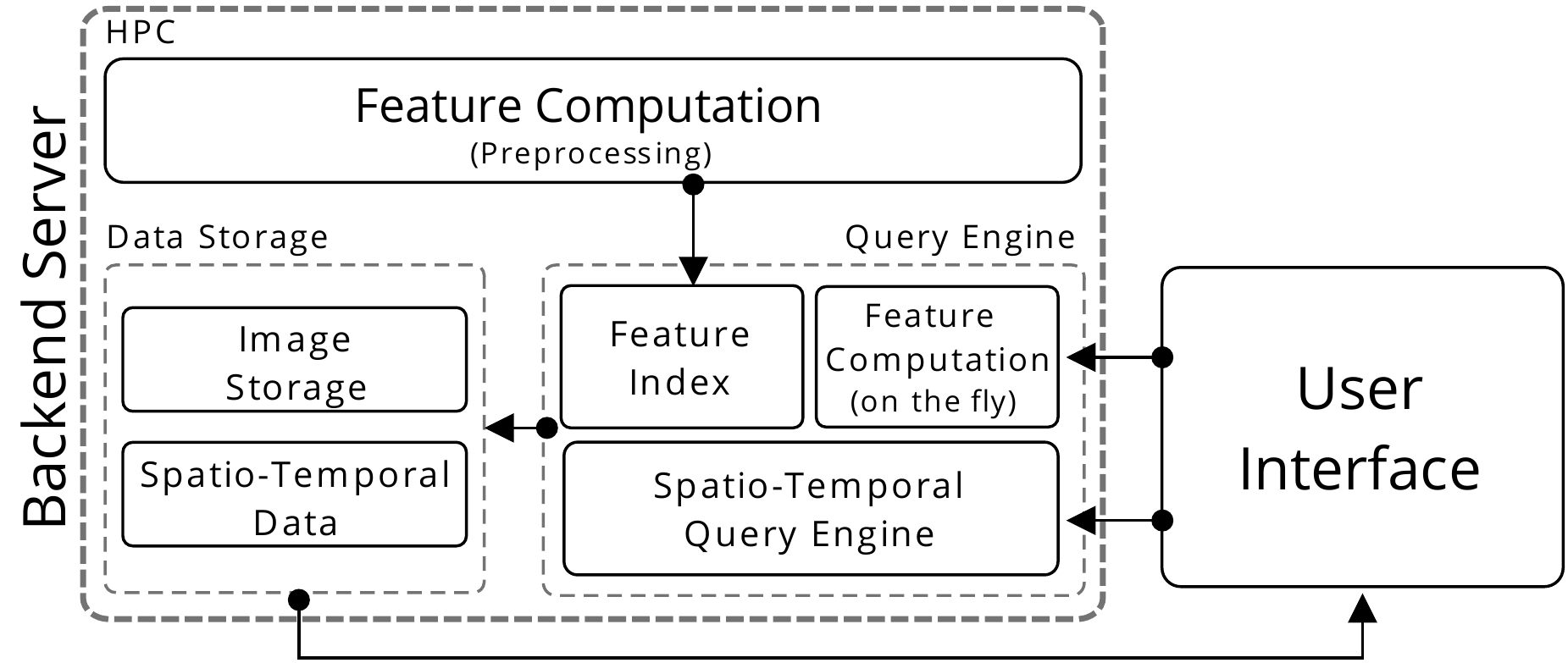}
	\caption{System architecture of the Urban Mosaic system.}
	\label{fig:system}
\end{figure}

\subsection{System Architecture}

Our description of \tool is broadly divided into two parts. First we discuss the backend server, which is responsible for enabling real-time responses to queries, followed by the user interface where these queries are composed~(Fig.~\ref{fig:system}).

\begin{figure*}[t]
	\centering
	\includegraphics[width=0.96\linewidth]{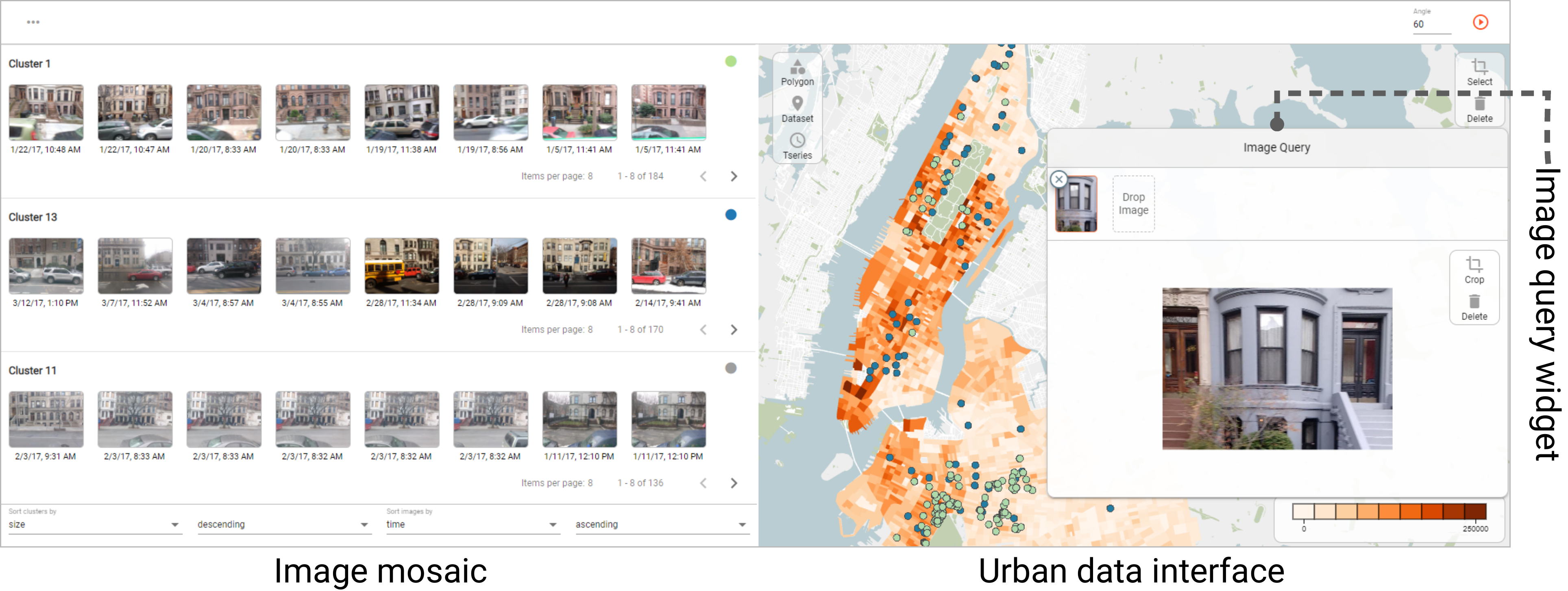}
	\caption{\tool user interface. A reference image is used in the image query widget to search the database for similar images. The image mosaic clusters the image query result and visualizes these clusters. The clusters, in this instance, are sorted based on their size. Two of the clusters are selected (blue and green circles) in the mosaic, showing the locations of the images part of these clusters on the map. Additionally, as part of the analysis, the median income data is visualized as a heatmap over the neighborhoods of NYC. One can notice that the locations of the images from the selected clusters are prominently concentrated in the affluent neighborhoods of NYC (relatively darker regions).
}
    \vspace{-0.2cm}
	\label{fig:interface}
\end{figure*}

\myparagraph{Query engine and data storage.}
The \tool query engine supports:
(1)~spatio-temporal selection and aggregation queries; and 
(2)~image similarity and clustering queries.
Spatio-temporal selection queries first perform a coarse temporal query to satisfy the time period constraint, and then the resulting coordinate points are tested in parallel against the spatial constraint. To handle spatio-temporal aggregation queries we again first run a coarse temporal query, which we pass to RasterJoin~\cite{zacharatou2017gpu} to compute the required spatial aggregation. For image similarity and clustering queries we preprocess a feature index to compute and store the images' features, and use this to perform similarity queries.
Images are stored as individual files on disk, while metadata and other spatio-temporal data are stored using MonetDBLite \cite{monetdblite-paper}.

\myparagraph{Image feature embedding.}
We use the intermediate layers of a pre-trained CNN as a representation of the visual information of an image. We adopt two approaches to achieve our goal of capturing semantic similarity: one for computing a coarse similarity measure for clustering queries, and another for the feature-based image similarity search. The first follows~\cite{tolias2016particular}, and derives a compact feature vector of size 512 from the convolutional layer activations of multiple regions of a single image. This produces a feature vector that coarsely captures \emph{general} aspects of the scene, to assess overall image similarity and group images into relevant clusters. The second, used for image similarity search, follows~\cite{sharif2014cnn} and employs a fully-connected layer of VGG16~\cite{simonyan2014very} as a compact and fixed-length feature descriptor of a image or region of a image. We compute a feature vector of size 4096 for each image region in a $2x2$ and a $4x4$ grid. Each image is represented by 20 4096-sized feature vectors in total. Two images are said to be similar if the similarity measure between at least one pair of feature vectors from the two images is below a given threshold. 

\begin{figure*}[t]
	\centering
	\includegraphics[width=0.96\linewidth]{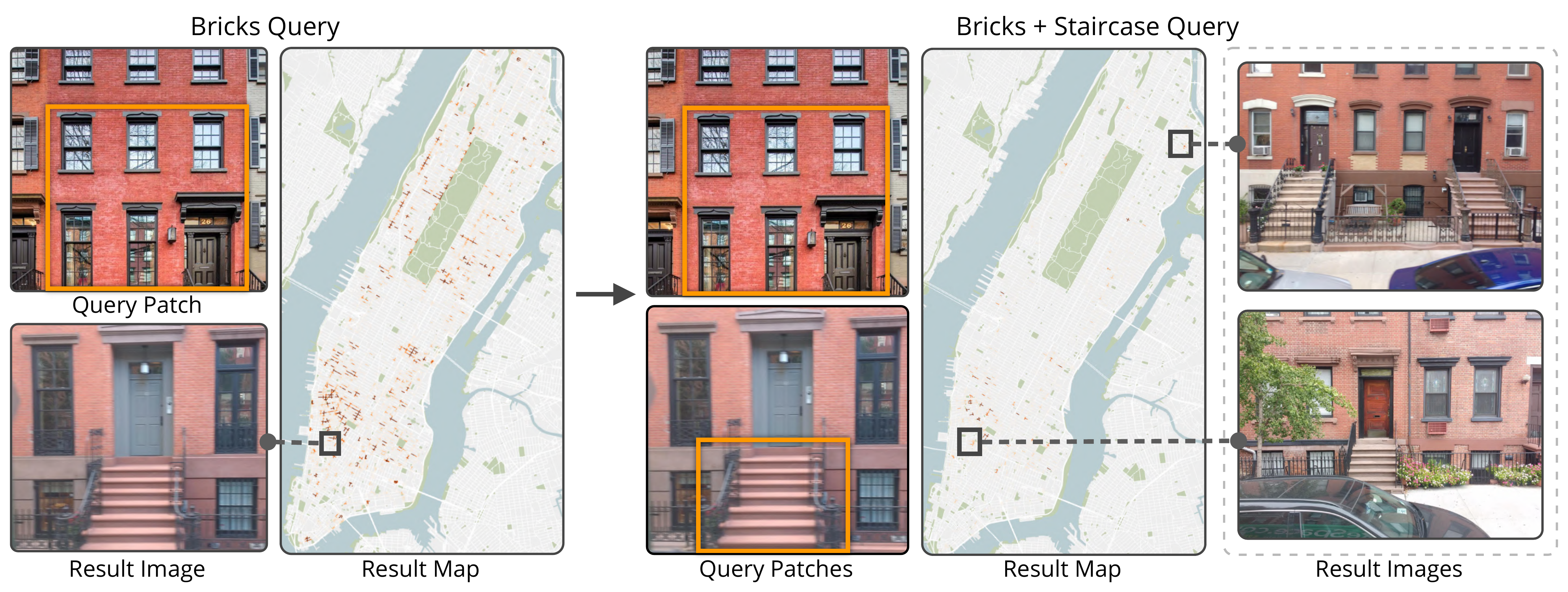}
	\caption{Composing image queries using \tool. A reference image of a red brick building is uploaded and cropped (orange outline) to search for red brick facades in NYC. The locations of the images with similar images are visualized on the map, and a region in lower Manhattan is then used to filter the image results. One such image shows a red brick facade with a staircase. The image query is now refined to add the staircase constraint. This is done by adding this image to the query, and cropping to the region containing the staircase. Two of the resulting images, containing both a red brick facade and a staircase, from different parts of Manhattan are shown on the right.
	}
	\vspace{-0.2cm}
	\label{fig:case02b}
\end{figure*}

\myparagraph{Efficient image similarity computation.}
We compute similarity between two images $I_1$ and $I_2$ as the angular distance $\alpha_{1,2}$ between the corresponding feature vectors $\vec{v_1}$ and $\vec{v_2}$:
\vspace{-0.15cm}
\[
\alpha_{1,2} = \cos^{-1}(\frac{\vec{v_1} \cdot \vec{v_2}}{\mid \vec{v_1} \mid\mid \vec{v_2} \mid})
\]
While this is straightforward for two images, a brute force search through all the feature vectors does not scale when working with several million images. 
The main challenges in enabling interactive similarity queries are twofold:
(1)~size: since a given image is composed of 20 4096-sized and one 512-sized feature vectors, it requires a total of 322~KB space. A data set consisting of 7.7M images will therefore require over 2.5~TB;
and (2)~complex floating point operations: a single comparison requires $O(4096)$ (or $O(512)$ for clustering queries) floating point operations, which while reasonable for a small number of comparisons, cannot be done interactively over the entire data.

We overcome this 
by trading off accuracy for speed using a locality sensitive hashing~(LSH)~\cite{Charikar2002} scheme to encode the feature vectors, and performing the query using the hashed data. 
%
Given a set of input vectors in $\Rspace^d$, the family of hash functions
making up this LSH scheme is defined as follows: choose a random $d$-dimensional vector $\vec{r}$
where each coordinate is drawn from a 1-dimensional Gaussian distribution. The hash function
corresponding to $\vec{r}$ is defined as:
\[
h_{\vec{r}}(\vec{u}) = 
\begin{cases}
1 & \text{if } \vec{r}\cdot \vec{u} \geq 0 \\
0 & \text{if } \vec{r}\cdot \vec{u} < 0 \\
\end{cases}
\]
Then, given two vectors $\vec{v_1}$ and $\vec{v_2}$ in $\Rspace^d$:
\[
\text{Pr}[h_{\vec{r}}(\vec{v_1}) = h_{\vec{r}}(\vec{v_2})] = 1 - \frac{\alpha_{1,2}}{\pi}
\]
The idea then is to estimate the above probability to compute the angular distance between any two vectors. To do this, we first generate $n$ $d$-dimensional
vectors $\{r_1, r_2, \ldots, r_n\}$ as described above, where $d$ = 4096 (or 512).
Given a $d$-dimensional vector $\vec{v}$, define the hash value $h(\vec{v})$ as follows:
\[
h(\vec{v}) = [h_{\vec{r_1}}(\vec{v}), h_{\vec{r_2}}(\vec{v}), \ldots, h_{\vec{r_n}}(\vec{v})]
\]
Then, the angle $\alpha_{1,2}$ between two vectors $v_1$ and $v_2$ can be estimated using $h(\vec{v_1})$ and
$h(\vec{v_2})$ as follows:
\[
\alpha_{1,2} = (1 - Pr) \times \pi
\text{ , where }
Pr = \frac{\sum_{i=1}^{n} \overline{h_{\vec{r_i}}(\vec{v_1}) \oplus h_{\vec{r_i}}(\vec{v_2})}}{n}
\]
%
Note that since $h()$ is a $0/1$ vector, each dimension of this vector can be stored 
using a single bit. In other words, the space required to store a single hash value is just $n$ bits, thus reducing the space required to store the hashed feature vectors of our data to 21~GB (using $n=1024$) instead of the over 2.5~TB required for storing the original feature vectors.
Moreover, the main operation in 
the above equation is the XNOR operation over $n$ bits, which is available as
an intrinsic operation on current CPUs making this computation extremely efficient.

\subsection{User Interface}

The \tool user interface consists of three main components, 
shown in Fig.~\ref{fig:interface}. We discuss each of these in turn.


\myparagraph{Image query widget.}
This widget allows users to compose image query constraints. Users can either upload images of interest or use the images already in the underlying database. The widget also allows users to  optionally crop query images. When multiple images are part of the query constraint, the query returns images that satisfy all the constraints (i.e., an intersection operation is performed).
Fig.~\ref{fig:case02b} illustrates an example of composing image queries using \tool.

\myparagraph{Urban data interface.}
The urban data interface allows users to visualize the spatial and temporal distributions of urban data sets using a \textit{map} and a \textit{time series} widget respectively. Based on their exploration of these urban data sets, users can select regions of interest over the map, and temporal ranges using the time series, to act as additional constraints on the image query. Urban data sets can be visualized as heatmaps at different spatial resolutions, e.g., neighbourhood, block, street or as a high resolution grid. Users can optionally apply a custom polygonal partition of the city to provide spatial resolution, by uploading polygonal data. The map may also be used to visualize the results from the image query as a heatmap.

\myparagraph{Image mosaic.}
The image mosaic widget is used to display clusters of similar images returned from image queries. Users can sort clusters, and images within a cluster, using data attributes from other urban data sets that are loaded into \tool (\eg noise complaints, temperature) or by image attributes from metadata (\eg image date, car id). Users can also visualize the spatial distribution of one or more selected clusters on the map, enabling them to more quickly identify locations where images are found with similar attributes to those in the selected cluster(s). Image mosaic is paginated to allow for queries that return large numbers of clusters.
A workspace widget provides a space for the user to save the potentially large number of images of interest that might result from multiple queries over the data undertaken during a session. These saved images can be exported, along with any other urban data associated with them during analysis.

\subsubsection{Practitioner perspective}
\highlight{
In an earlier prototype of \tool, practitioners found selecting between resolutions (e.g., $2x2$ or $4x4$ grid) when performing similarity search unnecessarily complex, and so in the prototype reported here the search is performed on all feature vectors, irrespective of its image region. Another feature introduced as a response to practitioner feedback is the facility to aggregate spatial urban data on multiple levels (e.g., neighbourhood, census tract). This was implemented because urban planners from Draw Brooklyn were particularly interested in analyzing the distribution of historical buildings at a census tract level.
}
\section{Using Urban Mosaic}

Urban Mosaic has been developed as a collaboration between researchers in visual analytics, HCI, and urban planning. However, to better understand the requirements practitioners might have, and the use cases this type of tool might facilitate, we undertook a series of interviews with practitioners from KPF, Draw Brooklyn and a research occupational therapist from NYU who specializes in urban accessibility for older people.
%
In this section we provide a detailed presentation of  two of the use cases that emerged from these interviews. In the first, we help identify locations were accessibility solutions for some may not always mean increased accessibility for all, and in the second we show how sensitivity towards preserving the urban fabric should extend beyond neighborhoods considered historic.
\highlight{All figures in this section are \tool screenshots edited for privacy, and to provide a clearer graphic explanation of the tasks undertaken.}

\subsection{Accessibility and Mobility}
Sidewalks are arguably the most important pedestrian-dedicated planned public spaces; and so inclusive, accessible streets, which serve a variety of users, are a mark of how effective a city is at the human scale. Title II of the Americans with Disabilities Act (ADA), requires pedestrian crossings to be accessible to people with disabilities~\cite{adatitle2}, and one challenge experienced by wheelchair users is coping with raised curbs. This can be mitigated with the addition of a short ramp cutting through the curb. Among the regulations guiding the use of curb ramps, is the requirement that a ``detectable warning", in the form of a tactile square, be installed on all the ramp-cuts for visually impaired pedestrians. These tactile pavings are distinctive, often yellow or red, with a bumpy surface created to be detectable with cane, feet or by guide dogs, and a bar pattern that serves as a directional guide.

In a 2014 study, the installation of these tactile pavings ranked as the least observed curb-cut regulation in Manhattan, with more than 88 percent of the curbs having no detectable warning surface installed ~\cite{accessible}. However, following legal action the city has agreed to upgrade all curbs to meet ADA ramp and tactile paving requirements~\cite{lawsuit}. This requires an assessment of the condition and quality of all the city's 162,000 curbs. The size of the task ahead becomes apparent when we see, by way of comparison, that 40 auditors were trained and employed for the 2014 study, in which less than 1\% of curbs were assessed. Here we describe how Urban Mosaic can help with this task.

\begin{figure}[!t]
 	\centering
 	\includegraphics[width=0.975\linewidth]{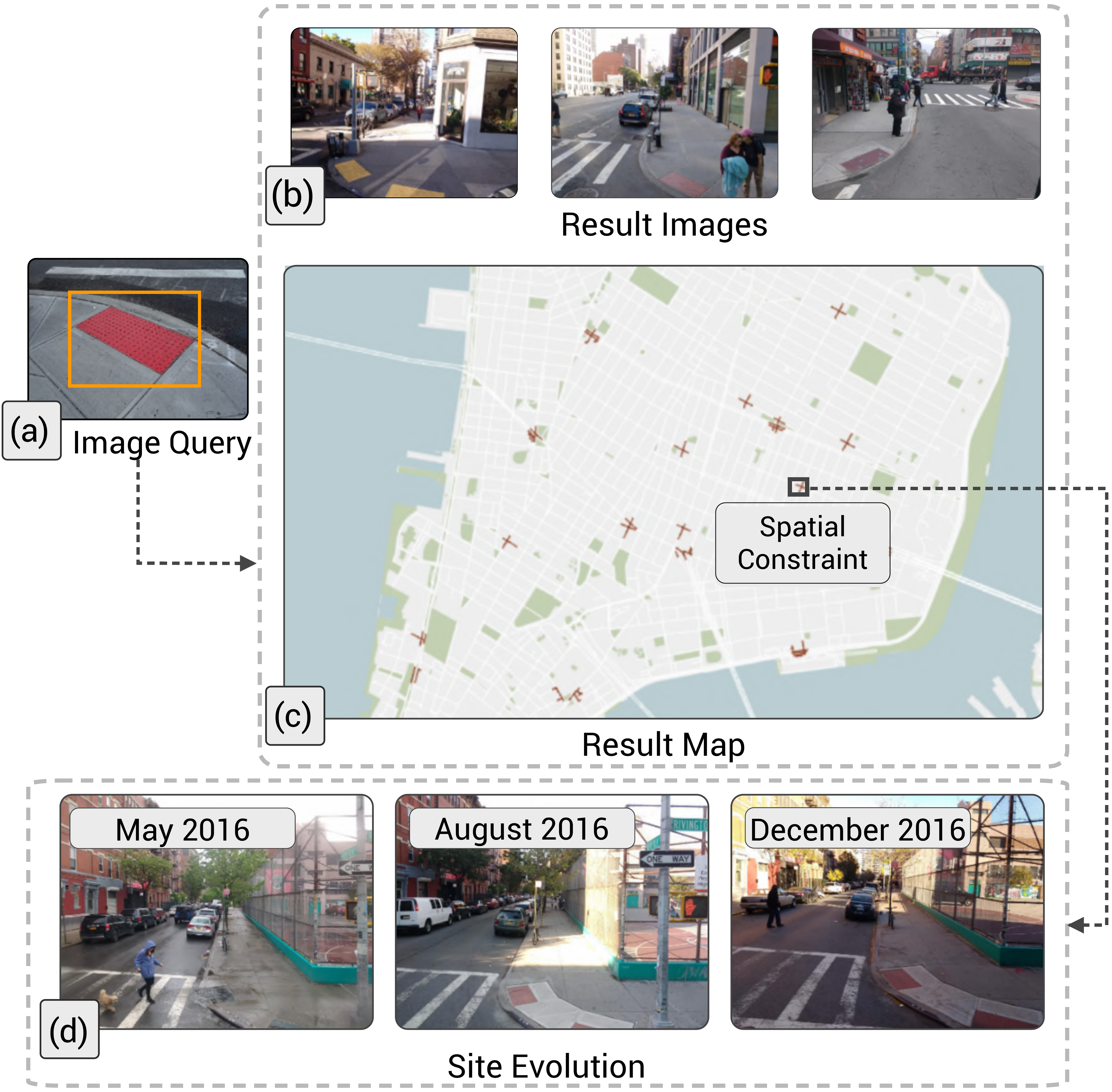}
 	\caption{
 	Using Urban Mosaic to inspect the presence and condition of tactile pavements in NYC, which is important to provide safe pedestrian access to people with movement or vision impairment disabilities.
\textbf{(a)}~Query image.
\textbf{(b)}~Three examples of other images in the database similar to the query image.
\textbf{(c)}~Visualizing the density of all images in NYC similar to the the query image as a heatmap. The highlighted region is set as a spatial constraint to query only for images similar to the query image in that region.
\textbf{(d)}~Three example images taken at different times showing that the pavement at that location is in dire need of repairs in May 2016, and was later repaired before August 2016.
 	}
	\label{fig:case01a}
\end{figure}

\myparagraph{Tracing the installation of tactile pavings.}
The first step in our assessment analysis is to select example images of tactile pavings installed on city curb ramps. We use a reference image (Fig.~\ref{fig:case01a}(a)) to provide the initial query input for our search, which will return clusters of similar images from the data set. These images are tagged with geo-location metadata, and so we are able to visualize the locations where images of tactile pavings were found on the city map (Fig.~\ref{fig:case01a}(c)). It is also possible to map the street corners where no images of tactile pavings were found, but it is important to remember that the lack of an image does not necessarily mean there is no tactile paving installed on the curb ramp at that site. In some cases this may be because the location has not been sufficiently covered in the database of images, it may also be because image quality (e.g., poor focus) was not sufficiently good for the algorithm to clearly identify the feature. Urban Mosaic was used for visual confirmation and query refinement of the results, as shown in Fig.~\ref{fig:case01a}(b).

The next step is to manually select a particular location of interest on the map, where tactile pavings are present, in order to gain some insight into the condition of the tactile,
and how long the tactile pavings have been in place. The selection of a particular location provides a spatial constraint for filtering our query results (Fig.~\ref{fig:case01a}(c)). 
We then sort the resulting images by the time that the picture was taken (from the image metadata) to analyze the location over time.
Looking over the sorted images from the selected location, we are able to identify that in May 2016 there was a damaged tactile pavement that became potentially dangerous to pedestrians because of a puddle of water (Fig.~\ref{fig:case01a}(d)~top). However, the tactile paving was later repaired (Fig.~\ref{fig:case01a}(d)~middle), and remained in good condition until the end of the year (Fig.~\ref{fig:case01a}(d)~bottom).

\begin{figure*}[t]
    \centering
    \includegraphics[width=0.96\linewidth]{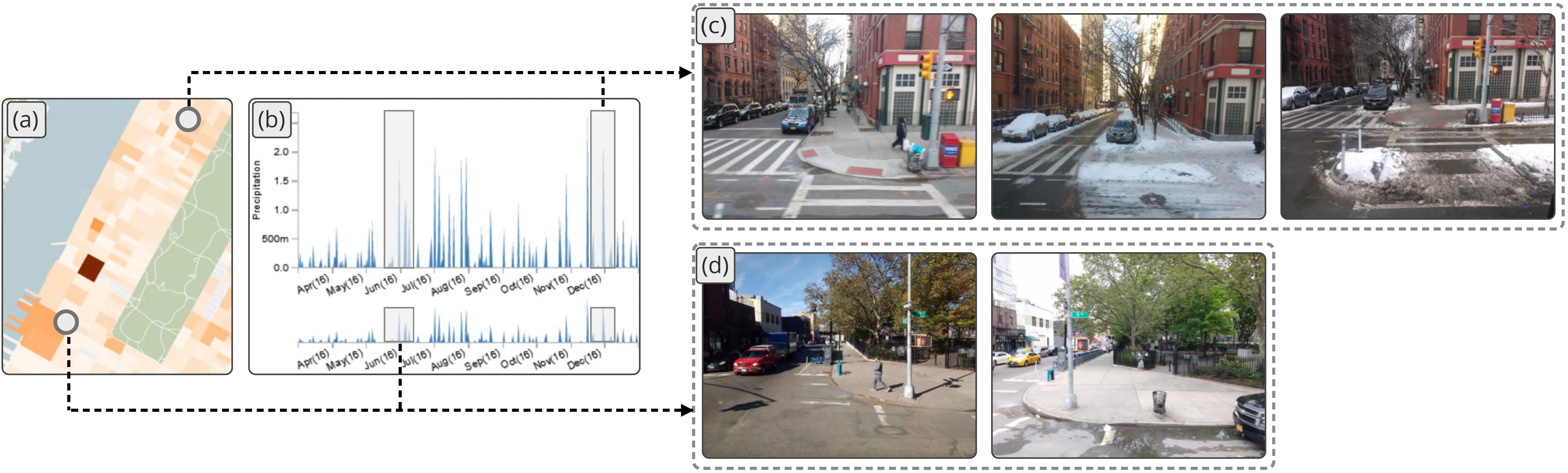}
    \caption{Assessing the conditions of pavements in regions having a higher concentration of older adults.
    (a)~Visualize the census data as a heatmap over the neighborhoods, and select spatial constraint based on locations with higher count of older adults.
    (b)~Select time periods having precipitation (rain or snow) as a temporal constraint. 
    (c)~A street corner in Upper West Side before and after snow fall.
    (d)~A street corner in Hell's Kitchen, having no tactile pavement, before and after rainfall.
    }
    \vspace{-0.2cm}
    \label{fig:empty-pavements}
\end{figure*}

\myparagraph{Tactile paving that may be a hazard for older adults.}
Our work with an occupational therapist began with an exploration of features in the urban fabric that might be associated with older adults falls.
%
%
We introduced our assessment analysis of tactile pavings as a possible model for finding something that might be helpful.
%
We learnt that the raised bumps in these tactile pavings are themselves often a cause of problems for older adults~\cite{doi:10.1680/muen.14.00016}, and that these problems are exacerbated in bad weather. With \tool we are able to combine data sets and image search to filter down areas of interest, and with the temporally dense data set of street-level images we can investigate the same location in different weather conditions. Next, we use \tool to identify neighborhoods with a larger population of older adults, and then to query how the curb ramp and tactile pavings look under different weather conditions in specific locations.

The first step in our extended assessment is to identify neighborhoods with a larger population of older adults. To do this we load and map NYC census data. We then zoom in on two neighborhoods: Hell's Kitchen (towards the bottom of the map in Fig.~\ref{fig:empty-pavements}(a)) and Upper West Side (towards the top). These provide a spatial constraint to our tactile paving image query, and we are able to identify examples from within each neighborhood. We then add a temporal constraint from weather data (Fig.~\ref{fig:empty-pavements}(b)), so that the query only returns images taken on days when there was some precipitation (rain and/or snow). 
We highlight three images from Upper West Side~(Fig.~\ref{fig:empty-pavements}(c)) which show the same street corner with different snow covering---no snow~(left), uncleared snow~(middle), and slushysnow~(right). Note that the two snow scenarios are particularly dangerous for older adults, increasing the likelihood of falls.
Similarly, we show two examples from Hell's Kitchen from another street corner (Fig.~\ref{fig:empty-pavements}(d)), without rain~(left) and after rain~(right). 
There are two interesting points to note in this example:
(1)~this corner does not have a tactile paving; and
(2)~the right image was taken after the rain stopped which can be seen by most of the area having dried out. However, the presence of puddles next to the curb indicates an area of concern for pedestrians, particularly older adults.

Because we are able to filter our image queries based on other spatial (e.g., census) and temporal (e.g., weather) data sets, we can present a detailed picture of particular locations of interest, based on the presence of tactile paving, and on a combination of demographic factors and weather conditions. This will allow the occupational therapist we are working with
to (1)~explore and identify unfamiliar neighborhoods that are causes of concern for older adults' pedestrian safety;
(2)~provide materials to inform discussions with residents, and plan specific strategies to support older adults' mobility and help prevent falls; and
(3)~use the examples to help create maps that can guide resilience training activities that must take place on good weather days, but which are sensitive to conditions on bad weather days.
This use case responds to a requirement to better understand the urban fabric in locations that may be geographically distant and unfamiliar, and to be sensitive to how the built environment in these locations are affected by different weather conditions. The ability to qualitatively reflect the temporal variation in a single location is particularly useful in this context.

\subsubsection{Practitioner perspective}

\highlight{
The ability to explore the same neighborhood under different weather conditions, in relatively fine-grained detail, supports a key task for our collaborating occupational therapist. Her practice involves developing resilience plans with older adults that aid mobility around their immediate neighborhood, and reduce the number and likelihood of falls. The program she runs is designed around in-field exploration and discussion of walking experiences. As the locations these sessions take place in can be almost anywhere across the city, she will rarely have a detailed knowledge of the particular hazards faced. However, there are potentially hazardous features that she will commonly look out for, such as a stoop without railings up the stairs, curb cuts where rainfall and snow might accumulate, areas with a risk of black ice, and tactile paving stones that can be problematic to people who use walking supports. As she explained during a requirements discussion, \emph{``It would be great if I was able to select neighborhoods where there are high concentration of seniors as well as poor street conditions, from broken sidewalks to accumulation of rainwater, and then do a comparative analysis between different neighborhoods.''}
  
She envisioned \tool as a tool to support route planning, which is an important aspect of planning for resilience training. Combining collections of images taken under different weather conditions with other types of data, such as the pattern of shadow at different times of the year, can help identify locations that are likely to be particularly prone to black ice. As she explained, \emph{``If seniors know the areas with higher risk of fall in advance, like those that get very little sunlight or much shade in the winter which can be the potential hot-spots for black ice, they can plan their walk to avoid those.''} During our formative evaluations she further explained that, \emph{``Looking at images from snowy days, I can evaluate the areas that are going to be the most problematic, like this segment or cross walk. If there is going to be the accumulation of water or snow then this area is a particularly bad one''}. Moreover, while there are a set of common mobility challenges older adults face, each different neighborhood is also likely to vary in important ways. Our aim with \tool is to support virtual pre-visits that can help improve understanding about the particular hazards faced, \emph{``Every neighborhood that we choose has its own challenges and what I like about the tool is that you can define what the main problem is, e.g inclement weather or obstruction, and these are the issues we're going to help the seniors identify and be safe around''}.

However, something we learned as our discussions progressed was that for her program, the best way to use the information that \tool can generate is typically using printed maps, and these maps should present information in a simple manner so that they can be a prompt for discussion and training. During an evaluation session we were told, \emph{``When you go to the meeting centers, some of them don't even have WiFi available or a computer screen, everything is very low-tech. I print out the maps, I take a highlighter and I show them: here is your block, these are the problematic locations.  So something I can translate to a hard copy that can be handed out is what works best at this stage''}. We were asked, \emph{``Is there a way to present that data in the form of a map, can you generate the hard copy of the map with the locations indicated on it?''}, and instructed that \emph{``the more user friendly the better and the more simplistic we can make them the better, because then it is going to be beneficial to everybody, not just people with higher education''}. This is a focus for ongoing work.

}

\begin{figure*}[!t]
 	\centering
 	\includegraphics[width=0.96\linewidth]{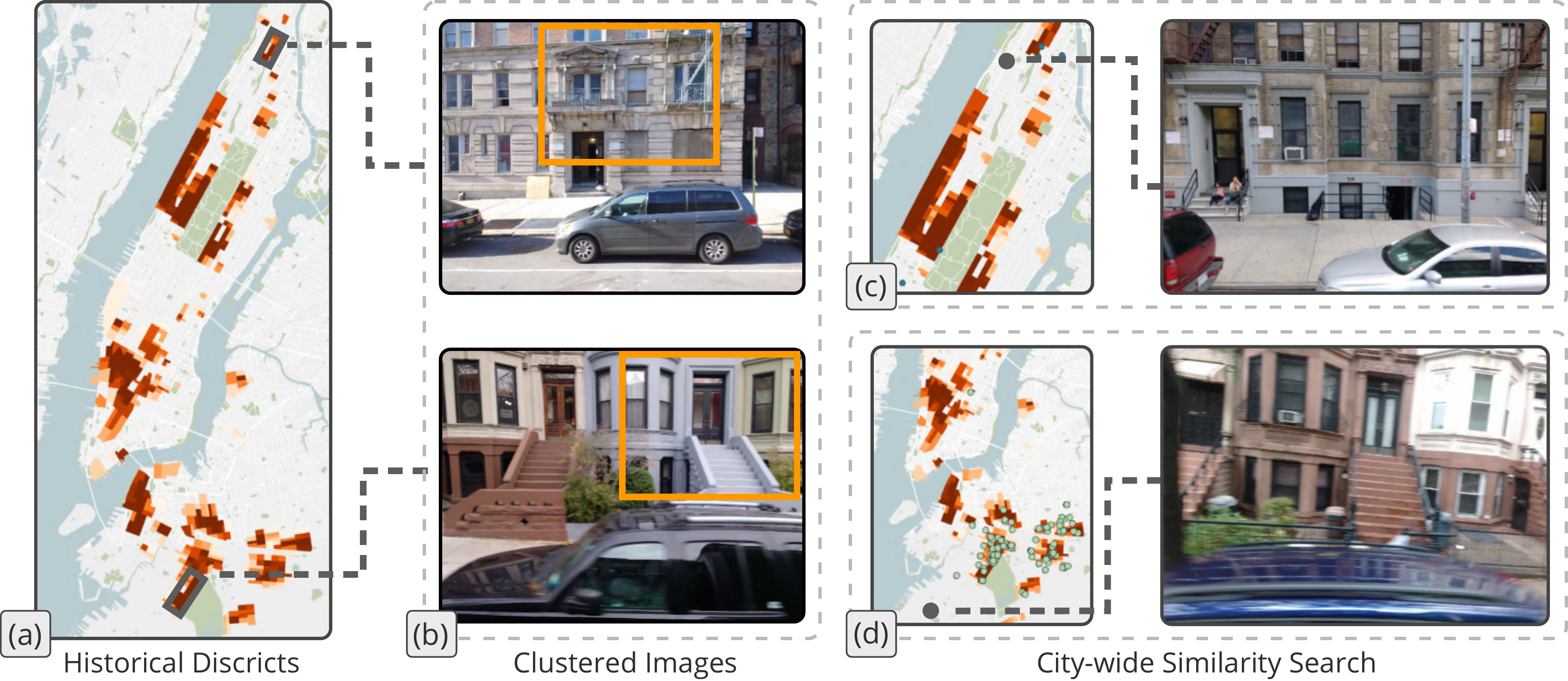}
 	\caption{Exploring the architectural characteristics of different neighborhoods. 
 	(a)~Heatmap showing the density of historical buildings in each neighborhood.
 	(b)~Example images showing the typical architectural characteristic of two historical districts: Hamilton~Heights~(top) and Park~Slope~(bottom). 
 	(c)~Searching for other regions in the city having the specific characteristic (orange region in (b).top) results in identifying a region in West~Harlem having similar characteristics.
 	Left: blocks designated as historical in NYC. Top right: most common types of arc.
 	(d)~Similarly, Sunset Park has characteristics similar to that of Park~Slope.
    }
    \vspace{-0.25cm}
	\label{fig:case02a}
\end{figure*}

\subsection{Preservation and Retrofitting}
In a large and dynamic city such as NYC, demands for new development are often in conflict with a desire to protect the particular character of individual neighborhoods, with Jane Jacobs' fight to protect Greenwich Village in the face of Robert Moses' West Village Urban Renewal and the Lower Manhattan Expressway project being the iconic example~\cite{gratz2010battle, jacobs_death_1961}. The different architectural styles present across the city reflect and define the rhythm of different neighborhoods. For example, the design of building facades has significant impact on pedestrians' walking experiences. The fine detail and vertical articulation of facades, the presence of narrow units, and display windows, all contribute to making a walk seem shorter and less tiring; and such features lead to more inviting and walkable neighborhoods, helping to create a livelier city~\cite{gehl2013cities}. Reflecting this, the regulation codes for construction, renovation, and repair, can differ based on neighborhood characteristics, with some districts designated historic, or some individual buildings designated landmarks \cite{nycpreservation, gvshp20yrs}. However, the presence of these positive architectural features is arguably even more important in diverse locations beyond the historic neighborhoods where regulation helps protect them; and yet it is precisely these locations where the risk of unsympathetic development is greatest. Protecting the urban fabric in neighborhoods not already designated historic is an important and challenging task, made more difficult by the geographic spread of the neighborhoods in question. Here we demonstrate how Urban Mosaic can support these activities by collating examples of particular architectural features, and mapping their density across geographically distant locations.

\myparagraph{Assessing urban fabric preservation.}
NYC neighborhoods are often characterized by particular architectural features and materials; for example the walk-up steps and stoop common among Brooklyn brownstones, and the limestone used to build many Beaux-Arts style Manhattan townhouses.
The first step in our assessment analysis is to load data containing locations of historic buildings in NYC, visualizing it as a heatmap over the neighborhoods of the city (Fig.~\ref{fig:case02a}(a)).
We then select two neighborhoods that are designated historic
, Hamilton Heights in Manhattan and Park Slope in Brooklyn. We then query the most common images found, and select buildings that display typical characteristics of each neighborhood (Fig.~\ref{fig:case02a}(b)). 

The next step is to identify neighborhoods where buildings with these positive architectural features are present in a beneficial density, but which have not been designated historic. 
Buildings with these features are typically desirable, and locations where they are found are often associated with gentrification and rapid redevelopment. Using \tool, we can identify such neighborhoods by selecting salient regions from images exemplifying the architectural features of interest, and using these as input to city-wide image queries (one for each feature of interest, highlighted by the orange rectangles in Fig.~\ref{fig:case02a}(b)). This identifies a neighborhood in West Harlem with a number of buildings with similar features to those expected in a Manhattan limestone townhouse (highlighted by the black circle on the map in Fig.~\ref{fig:case02a}(c)), and a neighborhood in Sunset Park with a number of buildings with similar features to those expected in a Brooklyn brownstone (Fig.~\ref{fig:case02a}(d)). 
%
Without the regulatory protections afforded by the designation of historic neighborhood status, there is an increased risk that redevelopment will include construction that is unsympathetic to such neighborhood's particular character, and to the lived-experience of residents. 
This danger may be increased in neighborhoods that are home to underrepresented communities or with low socioeconomic indicators. Because of this, our analysis may be furthered using Urban Mosaic to compare neighborhoods by demographic and other indicators, e.g. age, income, employment, education, housing costs, etc. If, for example, we compare the median income of Sunset Park with that of Park Slope, we see that income in the neighborhood designated historic is markedly higher. This use case responds to a requirement for monitoring preservation and retrofitting in a context where gentrification and excessive redevelopment is often a pressing issue of concern for many residents of the neighborhoods affected. This type of query was raised by both urban design practices as important.

\subsubsection{Practitioner perspective}
\highlight{
For practitioners such as our collaborators at KPF and Draw~Brooklyn, the challenge is often to make comparisons between multiple different neighborhoods, which may be geographically distant, along a variety of different parameters. Both pushed us to further develop the capacity to include additional data sets so that, for example, the visual experience associated with architectural features might be considered in relation to particular demographic or zoning characteristics of a neighborhood. For KPF, \emph{``this can bring analytics into the experience and feeling of a neighbourhood''} and help planners \emph{``visually assess the characteristics of the neighbourhood and find those common to that specific zone type''}, which might help in planning and designing new neighborhood regeneration projects as \emph{``planners can first select the features they think visually describe that neighborhood and assess how common these features are within that region and also in another neighbourhood''}. Image clusters \emph{``can describe the visual characteristics in a faster way"}, and exploring temporally dense image data might enable planners \emph{``to see how the neighborhood fabric is deteriorating through time''}.

Similarly, for Draw Brooklyn, the ability to query on visible architectural characteristics and compare other factors, such as zoning and demographics was a way to consider the varying impact of initiatives such as Active Design~\cite{aia2007active}. They wanted to be able to explore the building wall plane across neighborhoods, \emph{``where the sidewalk meets the private property line''}, as this is important for preserving characteristics that contribute to the active feeling of a neighborhood, and also for planning features that enhance an active experience. For example, open-front stores and transparency change peoples' perception of distance such that, \emph{``a walk along 5th Avenue feels much shorter and much less tiring since the street canopy and architectural articulation of the facades of adjacent buildings are rich and varied. Sidewalks should be designed like football fields: people need something interesting to see every 10 yards"}. However, many of these features, such as the vertical rhythm of the buildings or buildings spaced closely together, are more prevalent in the same historical neighborhoods where gentrification and aggressive over-development can become a problem.

During evaluations, both KPF and Draw Brooklyn drew attention towards adding a natural language capacity to support the translation between textual queries and visual search.  As one practitioner from Draw Brooklyn put it, adding this feature \emph{``allows users to tag
images with keywords, then the user can create a new gallery based on this keyword''}. 
}

\section{Discussion}

\highlight{

The availability of geographically and temporally dense data sets of street-level images, combined with computer vision and visual search techniques, has the potential to offer powerful new tools to support the work of architects, and urban planners and designers. For example, facilitating new forms of virtual audit and assessment. While virtual auditing with publicly available tools such as GSV has started to explore this space and address some of the limitations of in-field auditing~\cite{rundle2011using, badland2009understanding, badland2010can, kelly2012using, charreire2014using}, three key challenges have been noted. First, these approaches have typically required trained experts to manually explore large collections of images ~\cite{phillips2017online, zhu2017reliability, bader2015development}; second, these collections of images, while geographically dense, are temporally sparse, making it difficult to track change over time~\cite{rundle2011using, bethlehem2014spotlight, ben2013virtual, phillips2017online}; and third, these approaches do not integrate external data that might filter image data or capture the additional features required by many audit tools~\cite{shatu2018development, wilson2012assessing, bethlehem2014spotlight}. Similar challenges have been noted when a crowdsourcing approach to accessibility auditing at scale was adopted~\cite{Saha:2019:PSW:3290605.3300292}. Even for practices using data-informed methods, assessing the qualities of different neighborhoods typically involves a manual, time-consuming process of documenting different locations, or a similarly time-consuming process of virtually walking through neighborhoods in GSV.

With \tool, we begin to address each of these challenges. First, we are able to provide an automated first pass visual search for features of interest, which can significantly reduce the overhead of manually searching through GSV and refine the geographical scale of virtual auditing. Second, we offer a dense temporal granularity that facilitates comparison of the same location over multiple points in time, supporting greater insight into the qualitative, human-scale impact on changes to the urban fabric; as illustrated in our case study, \emph{Tracing the installation of tactile paving}. Third, we offer practitioners the ability to incorporate additional spatio-temporal data sets, such as weather, demographics, zoning, property values, crime, etc. to further filter or inform visual search; as illustrated in our use cases  \emph{Assessing urban fabric preservation} and \emph{Tactile paving that may be a hazard for older adults}. Building on the capacities that are emerging through current virtual audit and assessment practices, \tool does not replace the need to visit selected locations, rather we aim to support practitioners' informed decision making by helping to filter and refine the selection of neighborhoods and locations to visit, and by providing new insights into temporal changes in the urban fabric. Our collaborating practitioners have to date been supportive of these attempts to  provide fast, efficient alternatives for many of the tasks conventionally done through manual, time consuming methods; a practitioner from Draw~Brooklyn commented that \emph{"it can dramatically transform the way cities are planned and operated"}.

In developing \tool, we have incorporated roughly 7.7~million images taken over the course of a year in two boroughs of NYC. To explore what amounts to several terabytes of data, we utilize computer vision techniques to extract image features. However, even working with computed features can be slow enough to hamper visual exploration~\cite{liu-heer@tvcg2014}, and so we use an LSH-based index to reduce the memory footprint of the workable data and allow for fast query response times.
}

\subsection{Limitations and future work}
\highlight{
This research is part of ongoing inquiry into the opportunities that new image data sets and new techniques for visual search offer practitioners such as urban planners and designers. The work we report here includes these practitioners in requirements gathering and formative evaluation activities. These interactions have highlighted a number of particular areas for future research, including the integration of a natural language tool to translate text input into visual search and providing printable output in the form of maps or other reports.  However, to more fully understand these opportunities, and their wider impact on the work of these professionals, future research should include longitudinal user evaluation studies with \tool being used in the day-to-day practice of our collaborators, and be expanded to include a wider diversity of possible future users.} 
\section{Conclusion}

In this paper we introduced \tool, a visual analysis tool for the interactive search and exploration of the urban fabric of NYC, 
and demonstrated the potential of large-scale street-level image data sets to support urban planning and design. We have described our work with professional practitioners to surface opportunities and requirements for employing these data, and shown how they might offer a route towards staying sympathetic to the impact at a human-scale caused by city-scale changes. We also presented two detailed use cases that demonstrate how \tool responds to these opportunities and requirements; and we have shown how the combination of geographically and temporally dense image data can be combined with other spatio-temporal urban data to provide a bridge between the city and the street.

\section{Acknowledgements}
We would like to thank Carmera for providing the NYC image data set that motivated this work. We also thank our colleagues from Kohn~Pedersen~Fox, Draw~Brooklyn, and New~York~University for their help in this research. This work was supported in part by: the Moore-Sloan Data Science Environment at NYU; NASA; NSF awards CNS-1229185, CCF-1533564, CNS-1544753, CNS-1730396, CNS-1828576, CNS-1626098; CNPq grant 305974/2018-1; FAPERJ grant E-26/202.915/2019; and the NVIDIA NVAIL at NYU. C.~T.~Silva is partially supported by the DARPA D3M program. Any opinions, findings, and conclusions or recommendations expressed in this material are those of the authors and do not necessarily reflect the views of DARPA. We also thank Nvidia Corporation for donating GPUs used in this research.

\balance{}

\bibliographystyle{SIGCHI-Reference-Format}
\bibliography{paper}


\begin{thebibliography}{00}


\ifx \showCODEN    \undefined \def \showCODEN     #1{\unskip}     \fi
\ifx \showDOI      \undefined \def \showDOI       #1{{\tt DOI:}\penalty0{#1}\ }
  \fi
\ifx \showISBNx    \undefined \def \showISBNx     #1{\unskip}     \fi
\ifx \showISBNxiii \undefined \def \showISBNxiii  #1{\unskip}     \fi
\ifx \showISSN     \undefined \def \showISSN      #1{\unskip}     \fi
\ifx \showLCCN     \undefined \def \showLCCN      #1{\unskip}     \fi
\ifx \shownote     \undefined \def \shownote      #1{#1}          \fi
\ifx \showarticletitle \undefined \def \showarticletitle #1{#1}   \fi
\ifx \showURL      \undefined \def \showURL       #1{#1}          \fi

\bibitem{AGHAABBASI2018475}
{Mahdi Aghaabbasi}, {Mehdi Moeinaddini}, {Muhammad~Zaly Shah}, {Zohreh
  Asadi-Shekari}, {and} {Mehdi~Arjomand Kermani}. 2018.
\newblock \showarticletitle{Evaluating the capability of walkability audit
  tools for assessing sidewalks}.
\newblock {\em Sustainable Cities and Society\/}  {37} (2018), 475 -- 484.
\newblock
\showISSN{2210-6707}


\bibitem{aia2007active}
{{American Institute of Architects New York Chapter}}. 2007.
\newblock {Fit-City: Promoting Physical Activity through Design}.
\newblock   (2007).
\newblock


\bibitem{4677356}
{Gennady {Andrienko}} {and} {Natalia {Andrienko}}. 2008.
\newblock \showarticletitle{Spatio-temporal aggregation for visual analysis of
  movements}. In {\em 2008 IEEE Symposium on Visual Analytics Science and
  Technology}. 51--58.
\newblock


\bibitem{anguelov2010google}
{Dragomir Anguelov}, {Carole Dulong}, {Daniel Filip}, {Christian Frueh},
  {St{\'e}phane Lafon}, {Richard Lyon}, {Abhijit Ogale}, {Luc Vincent}, {and}
  {Josh Weaver}. 2010.
\newblock \showarticletitle{{Google Street View}: Capturing the world at street
  level}.
\newblock {\em Computer\/} {43}, 6 (2010), 32--38.
\newblock


\bibitem{6875954}
{Sean~M. {Arietta}}, {Alexei~A. {Efros}}, {Ravi {Ramamoorthi}}, {and} {Maneesh
  {Agrawala}}. 2014.
\newblock \showarticletitle{City Forensics: Using Visual Elements to Predict
  Non-Visual City Attributes}.
\newblock {\em IEEE Transactions on Visualization and Computer Graphics\/}
  {20}, 12 (2014), 2624--2633.
\newblock


\bibitem{ariffin2013perceptions}
{Raja Noriza~Raja Ariffin} {and} {Rustam~Khairi Zahari}. 2013.
\newblock \showarticletitle{Perceptions of the urban walking environments}.
\newblock {\em Procedia-Social and Behavioral Sciences\/}  {105} (2013),
  589--597.
\newblock


\bibitem{babenko2014neural}
{Artem Babenko}, {Anton Slesarev}, {Alexandr Chigorin}, {and} {Victor
  Lempitsky}. 2014.
\newblock \showarticletitle{Neural codes for image retrieval}. In {\em European
  conference on computer vision}. Springer, 584--599.
\newblock


\bibitem{bader2015development}
{Michael~D.M. Bader}, {Stephen~J. Mooney}, {Yeon~Jin Lee}, {Daniel Sheehan},
  {Kathryn~M. Neckerman}, {Andrew~G. Rundle}, {and} {Julien~O. Teitler}. 2015.
\newblock \showarticletitle{Development and deployment of the Computer Assisted
  Neighborhood Visual Assessment System (CANVAS) to measure health-related
  neighborhood conditions}.
\newblock {\em Health \& place\/}  {31} (2015), 163--172.
\newblock


\bibitem{badland2010can}
{Hannah~M. Badland}, {Simon Opit}, {Karen Witten}, {Robin~A. Kearns}, {and}
  {Suzanne Mavoa}. 2010.
\newblock \showarticletitle{Can virtual streetscape audits reliably replace
  physical streetscape audits?}
\newblock {\em Journal of Urban Health\/} {87}, 6 (2010), 1007--1016.
\newblock


\bibitem{badland2009understanding}
{Hannah~M. Badland}, {Grant~M. Schofield}, {Karen Witten}, {Philip~J.
  Schluter}, {Suzanne Mavoa}, {Robin~A. Kearns}, {Erica~A. Hinckson}, {Melody
  Oliver}, {Hector Kaiwai}, {Victoria~G. Jensen}, {and} {others}. 2009.
\newblock \showarticletitle{Understanding the Relationship between Activity and
  Neighbourhoods (URBAN) Study: research design and methodology}.
\newblock {\em BMC Public Health\/} {9}, 1 (2009), 224.
\newblock


\bibitem{balali2015detection}
{Vahid Balali}, {Armin~Ashouri Rad}, {and} {Mani Golparvar-Fard}. 2015.
\newblock \showarticletitle{Detection, classification, and mapping of US
  traffic signs using {Google Street View} images for roadway inventory
  management}.
\newblock {\em Visualization in Engineering\/} {3}, 1 (2015), 15.
\newblock


\bibitem{batty2009cities}
{Michael Batty}. 2009.
\newblock \showarticletitle{Cities as Complex Systems: Scaling, Interaction,
  Networks, Dynamics and Urban Morphologies.}
\newblock  (2009).
\newblock


\bibitem{batty2013big}
{Michael Batty}. 2013.
\newblock \showarticletitle{Big data, smart cities and city planning}.
\newblock {\em Dialogues in Human Geography\/} {3}, 3 (2013), 274--279.
\newblock


\bibitem{becker2011tale}
{Richard~A. Becker}, {Ramon Caceres}, {Karrie Hanson}, {Ji~Meng Loh}, {Simon
  Urbanek}, {Alexander Varshavsky}, {and} {Chris Volinsky}. 2011.
\newblock \showarticletitle{A tale of one city: Using cellular network data for
  urban planning}.
\newblock {\em IEEE Pervasive Computing\/} {10}, 4 (2011), 18--26.
\newblock


\bibitem{ben2013virtual}
{Eran Ben-Joseph}, {Jae~Seung Lee}, {Ellen~K. Cromley}, {Francine Laden}, {and}
  {Philip~J. Troped}. 2013.
\newblock \showarticletitle{Virtual and actual: relative accuracy of on-site
  and web-based instruments in auditing the environment for physical activity}.
\newblock {\em Health \& place\/}  {19} (2013), 138--150.
\newblock


\bibitem{bethlehem2014spotlight}
{John~R. Bethlehem}, {Joreintje~D. Mackenbach}, {Maher Ben-Rebah}, {Sofie
  Compernolle}, {Ketevan Glonti}, {Helga B{\'a}rdos}, {Harry~R. Rutter},
  {H{\'e}l{\`e}ne Charreire}, {Jean-Michel Oppert}, {Johannes Brug}, {and}
  {Jeroen Lakerveld}. 2014.
\newblock \showarticletitle{The SPOTLIGHT virtual audit tool: a valid and
  reliable tool to assess obesogenic characteristics of the built environment}.
\newblock {\em International journal of health geographics\/} {13}, 1 (2014),
  52.
\newblock


\bibitem{bise2018sidewalks}
{R.~Devon Bise}, {John~C. Rodgers~III}, {Michael~A. Maguigan}, {Brian
  Beaulieu}, {William Keith}, {Chanda~L. Maguigan}, {and} {Qingmin Meng}. 2018.
\newblock \showarticletitle{Sidewalks as Measures of Infrastructure
  Inequities}.
\newblock {\em Southeastern Geographer\/} {58}, 1 (2018), 39--57.
\newblock


\bibitem{BROWNSON2009S99}
{Ross~C. Brownson}, {Christine~M. Hoehner}, {Kristen Day}, {Ann Forsyth}, {and}
  {James~F. Sallis}. 2009.
\newblock \showarticletitle{Measuring the Built Environment for Physical
  Activity: State of the Science}.
\newblock {\em American Journal of Preventive Medicine\/} {36}, 4, Supplement
  (2009), S99 -- S123.e12.
\newblock
\showISSN{0749-3797}


\bibitem{carrasco2015using}
{Roberto Carrasco-Hernandez}, {Andrew~R.D. Smedley}, {and} {Ann~R. Webb}. 2015.
\newblock \showarticletitle{Using urban canyon geometries obtained from {Google
  Street View} for atmospheric studies: Potential applications in the
  calculation of street level total shortwave irradiances}.
\newblock {\em Energy and Buildings\/}  {86} (2015), 340--348.
\newblock


\bibitem{Charikar2002}
{Moses~S. Charikar}. 2002.
\newblock \showarticletitle{Similarity Estimation Techniques from Rounding
  Algorithms}. In {\em Proceedings of the Thiry-fourth Annual ACM Symposium on
  Theory of Computing} {\em (STOC '02)}. ACM, New York, NY, USA, 380--388.
\newblock
\showISBNx{1-58113-495-9}


\bibitem{charreire2014using}
{H{\'e}l{\`e}ne Charreire}, {Joreintje~D Mackenbach}, {M Ouasti}, {Jeroen
  Lakerveld}, {Sofie Compernolle}, {M Ben-Rebah}, {Martin McKee}, {Johannes
  Brug}, {Harry Rutter}, {and} {J-M Oppert}. 2014.
\newblock \showarticletitle{Using remote sensing to define environmental
  characteristics related to physical activity and dietary behaviours: a
  systematic review (the SPOTLIGHT project)}.
\newblock {\em Health \& place\/}  {25} (2014), 1--9.
\newblock


\bibitem{CLIFTON200795}
{Kelly~J. Clifton}, {Andr\'ea D.~Livi Smith}, {and} {Daniel Rodriguez}. 2007.
\newblock \showarticletitle{The development and testing of an audit for the
  pedestrian environment}.
\newblock {\em Landscape and Urban Planning\/} {80}, 1 (2007), 95 -- 110.
\newblock
\showISSN{0169-2046}


\bibitem{1467360}
{Navneet {Dalal}} {and} {Bill {Triggs}}. 2005.
\newblock \showarticletitle{Histograms of oriented gradients for human
  detection}. In {\em 2005 IEEE Computer Society Conference on Computer Vision
  and Pattern Recognition}, Vol.~1. 886--893 vol. 1.
\newblock
\showISSN{1063-6919}


\bibitem{DiezRoux2003}
{Ana~V. Diez~Roux}. 2003.
\newblock \showarticletitle{Residential environments and cardiovascular risk}.
\newblock {\em Journal of Urban Health\/} {80}, 4 (2003), 569--589.
\newblock
\showISSN{1468-2869}


\bibitem{doersch:hal-01053876}
{Carl Doersch}, {Saurabh Singh}, {Abhinav Gupta}, {Josef Sivic}, {and} {Alexei
  Efros}. 2012.
\newblock \showarticletitle{What Makes Paris Look like Paris?}
\newblock {\em {ACM Transactions on Graphics}\/} {31}, 4 (2012).
\newblock


\bibitem{Doraiswamy:2018:IVE:3183713.3193559}
{Harish Doraiswamy}, {Eleni Tzirita~Zacharatou}, {Fabio Miranda}, {Marcos
  Lage}, {Anastasia Ailamaki}, {Cl\'{a}udio~T. Silva}, {and} {Juliana Freire}.
  2018.
\newblock \showarticletitle{Interactive Visual Exploration of Spatio-Temporal
  Urban Data Sets Using Urbane}. In {\em Proceedings of the 2018 International
  Conference on Management of Data}. 1693--1696.
\newblock


\bibitem{emery2003reliability}
{James Emery}, {Carolyn Crump}, {and} {Philip Bors}. 2003.
\newblock \showarticletitle{Reliability and validity of two instruments
  designed to assess the walking and bicycling suitability of sidewalks and
  roads}.
\newblock {\em American Journal of Health Promotion\/} {18}, 1 (2003), 38--46.
\newblock


\bibitem{10.2307/26203303}
{Marcos~A.G. Ferreira} {and} {Suely da Penha~Sanches}. 2007.
\newblock \showarticletitle{Proposal of a Sidewalk Accessibility Index}.
\newblock {\em Journal of Urban and Environmental Engineering\/} {1}, 1 (2007),
  1--9.
\newblock
\showISSN{19823932}


\bibitem{ferreira2013visual}
{Nivan Ferreira}, {Jorge Poco}, {Huy~T Vo}, {Juliana Freire}, {and}
  {Cl{\'a}udio~T. Silva}. 2013.
\newblock \showarticletitle{Visual exploration of big spatio-temporal urban
  data: A study of {New York City} taxi trips}.
\newblock {\em IEEE Transactions on Visualization and Computer Graphics\/}
  {19}, 12 (2013), 2149--2158.
\newblock


\bibitem{gvshp20yrs}
{{Greenwich Village Society} for Historic~Preservation}. 2017.
\newblock Twenty years of preserving Federal-era row-houses.
\newblock   (2017).
\newblock


\bibitem{doi:10.1177/0042098008093386}
{Ann Forsyth}, {Mary Hearst}, {J.~Michael Oakes}, {and} {Kathryn~H. Schmitz}.
  2008.
\newblock \showarticletitle{Design and Destinations: Factors Influencing
  Walking and Total Physical Activity}.
\newblock {\em Urban Studies\/} {45}, 9 (2008), 1973--1996.
\newblock


\bibitem{gehl2013cities}
{Jan Gehl}. 2013.
\newblock {\em Cities for people}.
\newblock Island press.
\newblock


\bibitem{glaeser2018big}
{Edward~L. Glaeser}, {Scott~Duke Kominers}, {Michael Luca}, {and} {Nikhil
  Naik}. 2018.
\newblock \showarticletitle{Big data and big cities: The promises and
  limitations of improved measures of urban life}.
\newblock {\em Economic Inquiry\/} {56}, 1 (2018), 114--137.
\newblock


\bibitem{lawsuit}
{{Gothamist}}. 2019.
\newblock {NYC Agrees To Make All Sidewalk Curbs Accessible To The Disabled }.
\newblock   (March 2019).
\newblock
\showURL{%
Retrieved August 15, 2019 from
  \url{http://gothamist.com/2019/03/21/sidewalk_curbs_accessible_nyc.php}}


\bibitem{gratz2010battle}
{Roberta~Brandes Gratz}. 2010.
\newblock {\em The Battle for Gotham: New York in the Shadow of Robert Moses
  and Jane Jacobs}.
\newblock Bold Type Books.
\newblock


\bibitem{hammon2015data}
{Mary Hammon}. 2015.
\newblock Data-Driven: Leveraging the potential of big data for planning.
\newblock   (April 2015).
\newblock
\showURL{%
Retrieved September 11, 2019 from
  \url{https://www.planning.org/planning/2015/apr/datadriven.htm}}


\bibitem{hara2014tohme}
{Kotaro Hara}, {Jin Sun}, {Robert Moore}, {David Jacobs}, {and} {Jon
  Froehlich}. 2014.
\newblock \showarticletitle{Tohme: detecting curb ramps in {Google Street View}
  using crowdsourcing, computer vision, and machine learning}. In {\em
  Proceedings of the 27th annual ACM symposium on User interface software and
  technology}. ACM, 189--204.
\newblock


\bibitem{harvey2014measuring}
{Chester~Wollaeger Harvey}. 2014.
\newblock \showarticletitle{Measuring streetscape design for livability using
  spatial data and methods}.
\newblock  (2014).
\newblock


\bibitem{Hoang-Vu:2014:TUR:2630729.2630746}
{Tuan-Anh Hoang-Vu}, {Vicki Been}, {Ingrid~Gould Ellen}, {Max Weselcouch},
  {and} {Juliana Freire}. 2014.
\newblock \showarticletitle{Towards Understanding Real-Estate Ownership in New
  York City: Opportunities and Challenges}. In {\em Proceedings of the
  International Workshop on Data Science for Macro-Modeling} {\em (DSMM'14)}.
  ACM, New York, NY, USA, Article 15, 15:1--15:2 pages.
\newblock
\showISBNx{978-1-4503-3012-1}


\bibitem{jacobs_death_1961}
{Jane Jacobs}. 1961.
\newblock {\em The Death and Life of Great American Cities}.
\newblock Random House, New York.
\newblock


\bibitem{jeffery1977crime}
{C.~Ray Jeffery}. 1977.
\newblock {\em Crime prevention through environmental design}.
\newblock Sage Publications Beverly Hills, CA.
\newblock


\bibitem{8047432}
{Alireza {Karduni}}, {Isaac {Cho}}, {Ginette {Wessel}}, {William {Ribarsky}},
  {Eric {Sauda}}, {and} {Wenwen {Dou}}. 2017.
\newblock \showarticletitle{Urban Space Explorer: A Visual Analytics System for
  Urban Planning}.
\newblock {\em IEEE Computer Graphics and Applications\/} {37}, 5 (2017),
  50--60.
\newblock


\bibitem{doi:10.1123/jpah.2018-0476}
{Peter~T. Katzmarzyk}, {Kara~D. Denstel}, {Kim Beals}, {Jordan Carlson},
  {Scott~E. Crouter}, {Thomas~L. McKenzie}, {Russell~R. Pate}, {Susan~B.
  Sisson}, {Amanda~E. Staiano}, {Heidi Stanish}, {Dianne~S. Ward}, {Melicia
  Whitt-Glover}, {and} {Carly Wright}. 2018.
\newblock \showarticletitle{Results from the United States 2018 Report Card on
  Physical Activity for Children and Youth}.
\newblock {\em Journal of Physical Activity and Health\/} {15}, S2 (2018),
  S422--S424.
\newblock


\bibitem{kelly2012using}
{Cheryl~M. Kelly}, {Jeffrey~S. Wilson}, {Elizabeth~A. Baker}, {Douglas~K.
  Miller}, {and} {Mario Schootman}. 2012.
\newblock \showarticletitle{Using {Google Street View} to audit the built
  environment: inter-rater reliability results}.
\newblock {\em Annals of Behavioral Medicine\/}  {45} (2012), S108--S112.
\newblock


\bibitem{kopf2010street}
{Johannes Kopf}, {Billy Chen}, {Richard Szeliski}, {and} {Michael Cohen}. 2010.
\newblock \showarticletitle{Street slide: browsing street level imagery}. In
  {\em ACM Transactions on Graphics}, Vol.~29. ACM, 96:1--96:8.
\newblock


\bibitem{lander2017inferring}
{Christian Lander}, {Frederik Wiehr}, {Nico Herbig}, {Antonio Kr{\"u}ger},
  {and} {Markus L{\"o}chtefeld}. 2017.
\newblock \showarticletitle{Inferring landmarks for pedestrian navigation from
  mobile eye-tracking data and {Google Street View}}. In {\em Proceedings of
  the 2017 CHI Conference Extended Abstracts on Human Factors in Computing
  Systems}. 2721--2729.
\newblock


\bibitem{laurence2006death}
{Peter~L. Laurence}. 2006.
\newblock \showarticletitle{The death and life of urban design: Jane Jacobs,
  The Rockefeller Foundation and the new research in urbanism, 1955--1965}.
\newblock {\em Journal of Urban Design\/} {11}, 2 (2006), 145--172.
\newblock


\bibitem{lecue2014star}
{Freddy L{\'e}cu{\'e}}, {Simone Tallevi-Diotallevi}, {Jer Hayes}, {Robert
  Tucker}, {Veli Bicer}, {Marco~Luca Sbodio}, {and} {Pierpaolo Tommasi}. 2014.
\newblock \showarticletitle{Star-city: semantic traffic analytics and reasoning
  for city}. In {\em Proceedings of the 19th international conference on
  Intelligent User Interfaces}. ACM, 179--188.
\newblock


\bibitem{li2017mapping}
{Xiaojiang Li}, {Carlo Ratti}, {and} {Ian Seiferling}. 2017.
\newblock \showarticletitle{Mapping urban landscapes along streets using
  {Google street view}}. In {\em International cartographic conference}.
  Springer, 341--356.
\newblock


\bibitem{li2016urban}
{Xiaojiang Li} {and} {Chuanrong Zhang}. 2016.
\newblock \showarticletitle{Urban Land Use Information Retrieval Based on Scene
  Classification of {Google Street View} Images.}. In {\em SDW@ GIScience}.
  41--46.
\newblock


\bibitem{li2015assessing}
{Xiaojiang Li}, {Chuanrong Zhang}, {Weidong Li}, {Robert Ricard}, {Qingyan
  Meng}, {and} {Weixing Zhang}. 2015.
\newblock \showarticletitle{Assessing street-level urban greenery using {Google
  Street View} and a modified green view index}.
\newblock {\em Urban Forestry \& Urban Greening\/} {14}, 3 (2015), 675--685.
\newblock


\bibitem{liu-heer@tvcg2014}
{Zhicheng Liu} {and} {Jeffrey Heer}. 2014.
\newblock \showarticletitle{The Effects of Interactive Latency on Exploratory
  Visual Analysis}.
\newblock {\em IEEE Transactions on Visualization and Computer Graphics\/}
  {20}, 12 (2014), 2122--2131.
\newblock


\bibitem{Lowe:2004:DIF:993451.996342}
{David~G. Lowe}. 2004.
\newblock \showarticletitle{Distinctive Image Features from Scale-Invariant
  Keypoints}.
\newblock {\em Int. J. Comput. Vision\/} {60}, 2 (2004), 91--110.
\newblock
\showISSN{0920-5691}


\bibitem{mahyar2018communitycrit}
{Narges Mahyar}, {Michael~R James}, {Michelle~M Ng}, {Reginald~A Wu}, {and}
  {Steven~P Dow}. 2018.
\newblock \showarticletitle{CommunityCrit: Inviting the Public to Improve and
  Evaluate Urban Design Ideas through Micro-Activities}. In {\em Proceedings of
  the 2018 CHI Conference on Human Factors in Computing Systems}. 195.
\newblock


\bibitem{accessible}
{{Manhattan Borough President}}. 2018.
\newblock {Accessible Manhattan: Making Sidewalks Safe and Navigable for All}.
\newblock   (2018).
\newblock


\bibitem{marshall2010effect}
{Wesley~E. Marshall} {and} {Norman~W. Garrick}. 2010.
\newblock \showarticletitle{Effect of street network design on walking and
  biking}.
\newblock {\em Transportation Research Record\/} {2198}, 1 (2010), 103--115.
\newblock


\bibitem{8283638}
{Fabio Miranda}, {Harish Doraiswamy}, {Marcos Lage}, {Luc Wilson}, {Mondrian
  Hsieh}, {and} {Claudio~T. Silva}. 2019.
\newblock \showarticletitle{Shadow Accrual Maps: Efficient Accumulation of
  City-Scale Shadows Over Time}.
\newblock {\em IEEE Transactions on Visualization and Computer Graphics\/}
  {25}, 3 (2019), 1559--1574.
\newblock
\showISSN{1077-2626}


\bibitem{miranda2016urban}
{Fabio Miranda}, {Harish Doraiswamy}, {Marcos Lage}, {Kai Zhao}, {Bruno
  Gon{\c{c}}alves}, {Luc Wilson}, {Mondrian Hsieh}, {and} {Cl{\'a}udio~T
  Silva}. 2016.
\newblock \showarticletitle{{Urban Pulse}: Capturing the rhythm of cities}.
\newblock {\em IEEE transactions on visualization and computer graphics\/}
  {23}, 1 (2016), 791--800.
\newblock


\bibitem{naik2014streetscore}
{Nikhil Naik}, {Jade Philipoom}, {Ramesh Raskar}, {and} {C{\'e}sar Hidalgo}.
  2014.
\newblock \showarticletitle{Streetscore-predicting the perceived safety of one
  million streetscapes}. In {\em Proceedings of the 2014 IEEE Conference on
  Computer Vision and Pattern Recognition Workshops}. 779--785.
\newblock


\bibitem{10.1257/aer.p20161030}
{Nikhil Naik}, {Ramesh Raskar}, {and} {C{\'e}sar~A. Hidalgo}. 2016.
\newblock \showarticletitle{Cities Are Physical Too: Using Computer Vision to
  Measure the Quality and Impact of Urban Appearance}.
\newblock {\em American Economic Review\/} {106}, 5 (2016), 128--32.
\newblock


\bibitem{nycpreservation}
{{New York City Landmarks Preservation Commission}}. 2013.
\newblock {Rules of the New York City Landmarks Preservation Commission, Title
  63, Rules of the City of New York }.
\newblock   (2013).
\newblock


\bibitem{newman1972defensible}
{Oscar Newman}. 1972.
\newblock {\em Defensible space}.
\newblock Macmillan New York.
\newblock


\bibitem{doi:10.1680/muen.14.00016}
{Marcus Ormerod}, {Rita Newton}, {Hamish MacLennan}, {Mohamed Faruk}, {Sibylle
  Thies}, {Laurence Kenney}, {David Howard}, {and} {Chris Nester}. 2015.
\newblock \showarticletitle{Older people's experiences of using tactile
  paving}.
\newblock {\em Proceedings of the Institution of Civil Engineers - Municipal
  Engineer\/} {168}, 1 (2015), 3--10.
\newblock


\bibitem{7390069}
{Thomas {Ortner}}, {Johannes {Sorger}}, {Harald {Steinlechner}}, {Gerd
  {Hesina}}, {Harald {Piringer}}, {and} {Eduard {Gr\"{o}ller}}. 2017.
\newblock \showarticletitle{Vis-A-Ware: Integrating Spatial and Non-Spatial
  Visualization for Visibility-Aware Urban Planning}.
\newblock {\em IEEE Transactions on Visualization and Computer Graphics\/}
  {23}, 2 (2017), 1139--1151.
\newblock
\showISSN{1077-2626}


\bibitem{Pang:2019:CED:3290605.3300571}
{Carolyn Pang}, {Rui Pan}, {Carman Neustaedter}, {and} {Kate Hennessy}. 2019.
\newblock \showarticletitle{City Explorer: The Design and Evaluation of a
  Location-Based Community Information System}. In {\em Proceedings of the 2019
  CHI Conference on Human Factors in Computing Systems}. 341:1--341:15.
\newblock


\bibitem{phillips2017online}
{Christine~B Phillips}, {Jessa~K Engelberg}, {Carrie~M Geremia}, {Wenfei Zhu},
  {Jonathan~M Kurka}, {Kelli~L Cain}, {James~F Sallis}, {Terry~L Conway}, {and}
  {Marc~A Adams}. 2017.
\newblock \showarticletitle{Online versus in-person comparison of Microscale
  Audit of Pedestrian Streetscapes (MAPS) assessments: reliability of alternate
  methods}.
\newblock {\em International journal of health geographics\/} {16}, 1 (2017),
  27.
\newblock


\bibitem{psyllidis2015platform}
{Achilleas Psyllidis}, {Alessandro Bozzon}, {Stefano Bocconi}, {and}
  {Christiaan~Titos Bolivar}. 2015.
\newblock \showarticletitle{A platform for urban analytics and semantic data
  integration in city planning}. In {\em International conference on
  computer-aided architectural design futures}. Springer, 21--36.
\newblock


\bibitem{PURCIEL2009457}
{Marnie Purciel}, {Kathryn~M. Neckerman}, {Gina~S. Lovasi}, {James~W. Quinn},
  {Christopher Weiss}, {Michael~D.M. Bader}, {Reid Ewing}, {and} {Andrew
  Rundle}. 2009.
\newblock \showarticletitle{Creating and validating GIS measures of urban
  design for health research}.
\newblock {\em Journal of Environmental Psychology\/} {29}, 4 (2009), 457 --
  466.
\newblock
\showISSN{0272-4944}


\bibitem{4376168}
{Huamin {Qu}}, {{Wing-Yi} {Chan}}, {Anbang {Xu}}, {{Kai-Lun} {Chung}},
  {{Kai-Hon} {Lau}}, {and} {Ping {Guo}}. 2007.
\newblock \showarticletitle{Visual Analysis of the Air Pollution Problem in
  Hong Kong}.
\newblock {\em IEEE Transactions on Visualization and Computer Graphics\/}
  {13}, 6 (2007), 1408--1415.
\newblock
\showISSN{1077-2626}


\bibitem{quercia2014mining}
{Daniele Quercia} {and} {Diego Saez}. 2014.
\newblock \showarticletitle{Mining urban deprivation from foursquare: Implicit
  crowdsourcing of city land use}.
\newblock {\em IEEE Pervasive Computing\/} {13}, 2 (2014), 30--36.
\newblock


\bibitem{monetdblite-paper}
{Mark Raasveldt} {and} {Hannes M{\"{u}}hleisen}. 2018.
\newblock \showarticletitle{MonetDBLite: An Embedded Analytical Database}.
\newblock {\em CoRR\/}  {abs/1805.08520} (2018).
\newblock


\bibitem{reades2007cellular}
{Jonathan Reades}, {Francesco Calabrese}, {Andres Sevtsuk}, {and} {Carlo
  Ratti}. 2007.
\newblock \showarticletitle{Cellular census: Explorations in urban data
  collection}.
\newblock {\em IEEE Pervasive computing\/} {6}, 3 (2007), 30--38.
\newblock


\bibitem{doi:10.2105/AJPH.93.9.1456}
{Richard~A. Retting}, {Susan~A. Ferguson}, {and} {Anne~T. McCartt}. 2003.
\newblock \showarticletitle{A Review of Evidence-Based Traffic Engineering
  Measures Designed to Reduce Pedestrian-Motor Vehicle Crashes}.
\newblock {\em American Journal of Public Health\/} {93}, 9 (2003), 1456--1463.
\newblock


\bibitem{rogers2013social}
{Shannon~H. Rogers}, {Kevin~H. Gardner}, {and} {Cynthia~H. Carlson}. 2013.
\newblock \showarticletitle{Social capital and walkability as social aspects of
  sustainability}.
\newblock {\em Sustainability\/} {5}, 8 (2013), 3473--3483.
\newblock


\bibitem{rundle2011using}
{Andrew~G. Rundle}, {Michael~D.M. Bader}, {Catherine~A. Richards}, {Kathryn~M.
  Neckerman}, {and} {Julien~O. Teitler}. 2011.
\newblock \showarticletitle{Using {Google Street View} to audit neighborhood
  environments}.
\newblock {\em American journal of preventive medicine\/} {40}, 1 (2011),
  94--100.
\newblock


\bibitem{Saha:2019:PSW:3290605.3300292}
{Manaswi Saha}, {Michael Saugstad}, {Hanuma~Teja Maddali}, {Aileen Zeng}, {Ryan
  Holland}, {Steven Bower}, {Aditya Dash}, {Sage Chen}, {Anthony Li}, {Kotaro
  Hara}, {and} {Jon Froehlich}. 2019.
\newblock \showarticletitle{Project Sidewalk: A Web-based Crowdsourcing Tool
  for Collecting Sidewalk Accessibility Data At Scale}. In {\em Proceedings of
  the 2019 CHI Conference on Human Factors in Computing Systems}. 62:1--62:14.
\newblock


\bibitem{sallis2012role}
{James~F. Sallis}, {Myron~F. Floyd}, {Daniel~A. Rodr{\'\i}guez}, {and}
  {Brian~E. Saelens}. 2012.
\newblock \showarticletitle{Role of built environments in physical activity,
  obesity, and cardiovascular disease}.
\newblock {\em Circulation\/} {125}, 5 (2012), 729--737.
\newblock


\bibitem{sharif2014cnn}
{Ali Sharif~Razavian}, {Hossein Azizpour}, {Josephine Sullivan}, {and} {Stefan
  Carlsson}. 2014.
\newblock \showarticletitle{{CNN} features off-the-shelf: an astounding
  baseline for recognition}. In {\em Proceedings of the 2014 IEEE Conference on
  Computer Vision and Pattern Recognition Workshops}. 806--813.
\newblock


\bibitem{shatu2018development}
{Farjana Shatu} {and} {Tan Yigitcanlar}. 2018.
\newblock \showarticletitle{Development and validity of a virtual street
  walkability audit tool for pedestrian route choice analysis-SWATCH}.
\newblock {\em Journal of transport geography\/}  {70} (2018), 148--160.
\newblock


\bibitem{shen2018streetvizor}
{Qiaomu Shen}, {Wei Zeng}, {Yu Ye}, {Stefan~M{\"u}ller Arisona}, {Simon
  Schubiger}, {Remo Burkhard}, {and} {Huamin Qu}. 2018.
\newblock \showarticletitle{StreetVizor: Visual Exploration of Human-Scale
  Urban Forms Based on Street Views}.
\newblock {\em IEEE Transactions on Visualization and Computer Graphics\/}
  {24}, 1 (2018), 1004--1013.
\newblock


\bibitem{simonyan2014very}
{Karen Simonyan} {and} {Andrew Zisserman}. 2015.
\newblock \showarticletitle{Very Deep Convolutional Networks for Large-Scale
  Image Recognition}. In {\em 3rd International Conference on Learning
  Representations}.
\newblock


\bibitem{tang2018measuring}
{Jingxian Tang} {and} {Ying Long}. 2018.
\newblock \showarticletitle{Measuring visual quality of street space and its
  temporal variation: Methodology and its application in the Hutong area in
  Beijing}.
\newblock {\em Landscape and Urban Planning\/} (2018).
\newblock


\bibitem{thornton2016disparities}
{Christina~M. Thornton}, {Terry~L. Conway}, {Kelli~L. Cain}, {Kavita~A.
  Gavand}, {Brian~E. Saelens}, {Lawrence~D. Frank}, {Carrie~M. Geremia}, {Karen
  Glanz}, {Abby~C. King}, {and} {James~F. Sallis}. 2016.
\newblock \showarticletitle{Disparities in pedestrian streetscape environments
  by income and race/ethnicity}.
\newblock {\em SSM-population health\/}  {2} (2016), 206--216.
\newblock


\bibitem{tolias2016particular}
{Giorgos Tolias}, {Ronan Sicre}, {and} {Herv{\'e} J{\'e}gou}. 2016.
\newblock \showarticletitle{Particular object retrieval with integral
  max-pooling of CNN activations}.
\newblock  (2016).
\newblock


\bibitem{underkoffler1999urp}
{John Underkoffler} {and} {Hiroshi Ishii}. 1999.
\newblock \showarticletitle{Urp: a luminous-tangible workbench for urban
  planning and design}. In {\em Proceedings of the SIGCHI conference on Human
  Factors in Computing Systems}. ACM, 386--393.
\newblock


\bibitem{adatitle2}
{{United States Department of Justice Civil Rights Division}}. 2010.
\newblock {Curb Ramps and Pedestrian Crossings Under Title II of the ADA}.
\newblock   (2010).
\newblock


\bibitem{white2009sitelens}
{Sean White} {and} {Steven Feiner}. 2009.
\newblock \showarticletitle{SiteLens: situated visualization techniques for
  urban site visits}. In {\em Proceedings of the SIGCHI conference on human
  factors in computing systems}. ACM, 1117--1120.
\newblock


\bibitem{doi:10.1177/1524839903260595}
{Joel~E. Williams}, {Martin Evans}, {Karen~A. Kirtland}, {Marlo~M. Cavnar},
  {Patricia~A. Sharpe}, {Matthew~J. Neet}, {and} {Annette Cook}. 2005.
\newblock \showarticletitle{Development and Use of a Tool for Assessing
  Sidewalk Maintenance as an Environmental Support of Physical Activity}.
\newblock {\em Health Promotion Practice\/} {6}, 1 (2005), 81--88.
\newblock


\bibitem{wilson2012assessing}
{Jeffrey~S Wilson}, {Cheryl~M Kelly}, {Mario Schootman}, {Elizabeth~A Baker},
  {Aniruddha Banerjee}, {Morgan Clennin}, {and} {Douglas~K Miller}. 2012.
\newblock \showarticletitle{Assessing the built environment using
  omnidirectional imagery}.
\newblock {\em American journal of preventive medicine\/} {42}, 2 (2012),
  193--199.
\newblock


\bibitem{ye2018measuring}
{Yu Ye}, {Daniel Richards}, {Yi Lu}, {Xiaoping Song}, {Yu Zhuang}, {Wei Zeng},
  {and} {Teng Zhong}. 2018.
\newblock \showarticletitle{Measuring daily accessed street greenery: A
  human-scale approach for informing better urban planning practices}.
\newblock {\em Landscape and Urban Planning\/} (2018).
\newblock


\bibitem{yin2016measuring}
{Li Yin} {and} {Zhenxin Wang}. 2016.
\newblock \showarticletitle{Measuring visual enclosure for street walkability:
  Using machine learning algorithms and {Google Street View} imagery}.
\newblock {\em Applied geography\/}  {76} (2016), 147--153.
\newblock


\bibitem{zacharatou2017gpu}
{Eleni~Tzirita Zacharatou}, {Harish Doraiswamy}, {Anastasia Ailamaki},
  {Cl{\'a}udio~T. Silva}, {and} {Juliana Freire}. 2017.
\newblock \showarticletitle{{GPU} rasterization for real-time spatial
  aggregation over arbitrary polygons}.
\newblock {\em Proceedings of the VLDB Endowment\/} {11}, 3 (2017), 352--365.
\newblock


\bibitem{6876029}
{Wei {Zeng}}, {{Chi-Wing} {Fu}}, {Stefan~M. {Arisona}}, {Alexander {Erath}},
  {and} {Huamin {Qu}}. 2014.
\newblock \showarticletitle{Visualizing Mobility of Public Transportation
  System}.
\newblock {\em IEEE Transactions on Visualization and Computer Graphics\/}
  {20}, 12 (2014), 1833--1842.
\newblock
\showISSN{1077-2626}


\bibitem{7506246}
{Yixian {Zheng}}, {Wenchao {Wu}}, {Yuanzhe {Chen}}, {Huamin {Qu}}, {and}
  {Lionel~M. {Ni}}. 2016.
\newblock \showarticletitle{Visual Analytics in Urban Computing: An Overview}.
\newblock {\em IEEE Transactions on Big Data\/} {2}, 3 (2016), 276--296.
\newblock
\showISSN{2332-7790}


\bibitem{zhou2017recent}
{Wengang Zhou}, {Houqiang Li}, {and} {Qi Tian}. 2017.
\newblock \showarticletitle{Recent Advance in Content-based Image Retrieval:
  {A} Literature Survey}.
\newblock {\em CoRR\/}  {abs/1706.06064} (2017).
\newblock


\bibitem{zhu2017reliability}
{Wenfei Zhu}, {Yuliang Sun}, {Jonathan Kurka}, {Carrie Geremia}, {Jessa~K
  Engelberg}, {Kelli Cain}, {Terry Conway}, {James~F Sallis}, {Steven~P
  Hooker}, {and} {Marc~A Adams}. 2017.
\newblock \showarticletitle{Reliability between online raters with varying
  familiarities of a region: Microscale Audit of Pedestrian Streetscapes
  (MAPS)}.
\newblock {\em Landscape and urban planning\/}  {167} (2017), 240--248.
\newblock


\end{thebibliography}

\end{document}